\documentclass[twocolumn,trackchanges]{aastex631}

\usepackage{amsmath,amssymb}
\usepackage{physics}
\usepackage{amsfonts}
\usepackage{mathtools}

\usepackage{etoolbox}
\usepackage[refpage]{nomencl}
\usepackage{epigraph}
\usepackage[normalem]{ulem}  

\providetoggle{nomsort}
\settoggle{nomsort}{true} 

\makeatletter
\iftoggle{nomsort}{%
    \let\old@@@nomenclature=\@@@nomenclature        
        \newcounter{@nomcount} \setcounter{@nomcount}{0}%
        \newcommand{\threedigits}[1]{\ifnum#1<100 0\two@digits{#1} \else \number#1\fi}
        \renewcommand\the@nomcount{\threedigits{\value{@nomcount}}}
        \def\@@@nomenclature[#1]#2#3{
          \addtocounter{@nomcount}{1}%
        \def\@tempa{#2}\def\@tempb{#3}%
          \protected@write\@nomenclaturefile{}%
          {\string\nomenclatureentry{\the@nomcount\nom@verb\@tempa @[{\nom@verb\@tempa}]%
          \begingroup\nom@verb\@tempb\protect\nomeqref{\theequation}%
          |nompageref}{\thepage}}%
          \endgroup
          \@esphack}%
      }{}
\makeatother

\newcommand{\mynomone}[3][section]{%
  \begingroup\edef\x{\endgroup
  \unexpanded{\nomenclature{#2}}%
    {\unexpanded{#3} \hspace*{\fill}  (\csname the#1\endcsname)}}\x}

\newcommand{\mynomtwo}[4][section]{%
  \begingroup\edef\x{\endgroup
  \unexpanded{\nomenclature[#2]{#3}}%
    {\unexpanded{#4} \hspace*{\fill}  (\csname the#1\endcsname)}}\x}

\renewcommand\nomgroup[1]{%
  \item[\bfseries
  \ifstrequal{#1}{A}{Acronyms}{%
  \ifstrequal{#1}{S}{Symbols}{%
  \ifstrequal{#1}{C}{Other Symbols}{}}}%
]}

\newcounter{logglabel}

\makenomenclature

\newcommand{\eq}[1]{Eq.~(\ref{#1})}

\shorttitle{Capture and Decimation of CBPs}
\shortauthors{Farhat \& Touma}
\graphicspath{{./}{figures/}}

\begin{document}
\title{Capture into Apsidal Resonance and \\
the Decimation of Planets around In-spiraling Binaries }

\author[0000-0001-7864-6627]{Mohammad Farhat}
\altaffiliation{Miller Fellow}

\affiliation{Department of Astronomy, University of California, Berkeley, Berkeley, CA 94720-3411, USA  }
\affiliation{Department of Earth and Planetary Science, University of California, Berkeley, Berkeley, CA 94720-4767, USA}

\author{Jihad  Touma}
\affiliation{Department of Physics, American University of Beirut
PO Box 11-0236, Riad El-Solh, Beirut 11097 2020, Lebanon }
\affiliation{Center for Advanced Mathematical Sciences, American University of Beirut, PO Box 11-0236, Riad El-Solh, Beirut 11097 2020, Lebanon}



\begin{abstract}
Transiting circumbinary planets (CBPs) are conspicuously rare, and entirely absent around stellar binaries with periods $\leq7~{\rm days}$. Here, we exploit  a secular resonance to stimulate the orbit of a CBP into strong, disruptive interactions with the host binary. The process requires no tertiary companion and is triggered when the general-relativistic precession of a tightening binary matches the Newtonian precession it induces in its companion planet. Adiabatic capture in this resonance sees the binary draining angular momentum from the CBP's orbit which grows steadily in eccentricity until destabilization, and eventual ejection or engulfment. We map this resonance in phase space, then investigate the dynamical outcomes of encounter in the course of tidally shrinking binaries. With the help of orbit averaged simulations of a suite of systems, we find that, around tightening binaries: eight out of ten CBPs encounter and are captured in the resonance; three out of four are `destroyed'; and survivors lurk on remote, low-transit-probability orbits. This suggests that the very process which forms tight binaries effectively clears the region where transiting CBPs could reside. 
\end{abstract}



\section{Introduction} 
\label{sec:intro}

Eclipsing binary stars (EBs) are propitious hunting grounds for transiting circumbinary planets \citep[CBPs hereafter; e.g.,][]{borucki2016kepler}. Their alignment with our line of sight, together with likely coplanar CBP formation and evolution, improves the odds of detection \citep[][]{martin2015circumbinary,kostov2023transiting}. There are complications of course with the binary’s dynamics contributing significant variations in transit timing, depth, and duration. At the same time, those same variations provide a unique observational signature which facilitates ruling out false positives and allows accurate determination of the planetary mass and radius \citep[see e.g.,][]{welsh2018two}.

Planet formation around binaries is not expected to be any less efficient than around single stars \citep[e.g.,][]{nelson2003evolution,thebault2015planet,bromley2015planet}. Circumbinary discs will be left with substantial mass as the binary clears a sizable inner cavity, thus slowing down if not altogether suppressing accretion onto the stars \citep[e.g.,][]{artymowicz1994dynamics,vartanyan2016tatooine,pierens2023three}. Planet formation close to the cavity’s boundary may face difficulties via both planetesimal accretion [due to high collisional velocities \citep[e.g.,][]{marzari2008planetesimal,paardekooper2012not}], and pebble accretion [due to hydrodynamic turbulence enhanced by inner disc instabilities \citep[e.g.,][]{papaloizou2005local,barker2014hydrodynamic,pierens2021vertical}]. It was argued that disc self-gravity can mitigate against collisional fragmentation, in axisymmetric discs for sure, and to a lesser extent in eccentric ones \citep[e.g.,][]{rafikov2014planet,silsbee2014planet,silsbee2015birth}. Suffice to say that, while formation 
close-in remains under discussion, planets have what it takes to form in the disc's outskirts and migrate inwards under the disc's torques \citep[e.g.,][]{nelson2003evolution,kley2012planet,kley2015evolution}.

All things considered, the occurrence rate of CBPs, particularly those detectable by Kepler, should be comparable to the occurrence rate of planets around single stars \citep[${\sim}10\%$;][]{martin2014planets,armstrong2014abundance}.  Thus far, some 14 transiting CBPs have been detected by Kepler and TESS, occupying the short-orbital period end of a population of 47 candidates. Given a total of ${\sim}~3000$ EBs observed by Kepler \citep[][]{kirk2016kepler}, one can safely speak of a dearth in CBP detection. The alarmingly miniscule harvest looks worst when limiting oneself to CBPs around short-period binaries as two thirds of the Kepler-detected EBs are in tight orbits with periods shorter than 7 days, while the shortest period binary observed to host a CBP is Kepler-47, with a binary orbital period of 7.45 days \citep[][]{orosz2012kepler,orosz2019discovery}. Important to note that similar such negligible occurrence rate was confirmed in a CoRoT data sample of ${\sim}2000$ EBs \citep[][]{klagyivik2017limits}.

One wonders whether these metrics are indicative of physical processes sculpting the expected distribution of CBPs, or merely the result of observational limitations/bias \citep[e.g.,][]{li2016uncovering}. A lack of transiting CBPs around long-period binaries is understandable: the circumbinary instability region extends to planetary periods ${\sim}8$ times that of  the binary  \citep[][]{dvorak1986critical,holman1999long}, with consequent decay of planetary transit probability. One would then expect observational bias to favor CBP transits around short-period binaries, making their scarcity even more intriguing.

Allowing for disruptive physical processes, several studies \citep[see e.g., ][]{hamers2016triple,munoz2015survival} investigated the dynamical fate of a CBP when allowing for the effect of a tertiary star, which forces the central binary to undergo Kozai-Lidov cycles then shrink under tidal friction \citep[][]{fabrycky2007shrinking}. Others considered CBPs in isolated binary systems and explored the dynamical fate of the planet: i-when migrating toward the binary and encountering mean motion resonances \citep[MMR; e.g.,][]{sutherland2019instabilities,martin2022running,gianuzzi2023circumbinary}, or ii- when allowing for the tidal and magnetic evolution of the binary itself \citep[][]{fleming2018lack}. More recently, \citet[][]{mogan2025concealing} investigated the likelihood of transits around tight binaries having decayed orbitally under tides, leaving the initially closer-in planets stranded on non-transiting orbits.

In this \textit{Letter}, we add an overlooked, and we believe critical, dynamical twist to ongoing attempts at explaining the CBP desert. We report on the sculpting of a typical CBP distribution by a secular resonance between the apsidal precession of the binary (general relativistic effects included) and that of the planet. We note that the resonance in question is natural around the birthplace of CBPs in isolated binaries (or binaries that are part of triple systems for that matter). We then show how the sweeping of the resonance in the course of typical evolutionary scenarios perturbs planetary orbits to a point that the occurrence rate of CBPs can be significantly reduced when compared to that of planets orbiting single stars. Prior studies of planet formation and evolution around tightening binaries should be revisited with this resonance explicitly taken into consideration.

\section{Resonance Forensics}

We consider CBPs which are initially dynamically stable under strong three body perturbations, i.e. far enough from the binary to be safe from mean motion resonances, and yet close enough to be prone to disruption by a novel resonance which operates over longer/secular timescales, in the presence of non-conservative evolutionary processes. In other words, we will consider regimes where dynamical processes of interest operate over timescales which are long enough for them to be captured by an orbit averaged treatment\footnote{Planets far enough for their orbital periods to couple to long/secular timescales in the binary’s evolution are potentially prone to disruption by semi-secular resonances, evection being an archetype of the latter in hierarchical systems \citep[][]{touma1998resonances,touma2015disruption}.}. We refer to the orbital elements of the planet using the subscript ``p", and to those of the stars using the subscripts ``A" and ``B", and ``AB" for the binary. 

The resonance in question obtains when the apse precession rate of the host binary matches that of the CBP, i.e. when
\begin{equation}\label{Resonance_condition}
    \dot{\varpi}_{\rm AB} = \dot{\varpi}_{\rm p},
\end{equation}
where $\dot{\varpi}_{\rm AB}$ is the binary's apse precession rate under the effects of Post-Newtonian (PN)/General Relativistic (GR) corrections, tides, and/or the Newtonian forcing of the CBP (which is absent when the planet is treated as a test particle; refer to Appendix  for details). In a single planet system, and at distances of interest to us, the planet's apsidal precession is mainly controlled by the Newtonian forcing of the binary (see below for variations on this basic scenario). In this preliminary encounter with the resonance, we treat the planet as a test particle, relaxing this assumption in sections to follow. We further restrict to coplanar configurations all through, while commenting on spatial dynamics in the discussion. 

\begin{figure}[t]
\includegraphics[width=.47\textwidth]{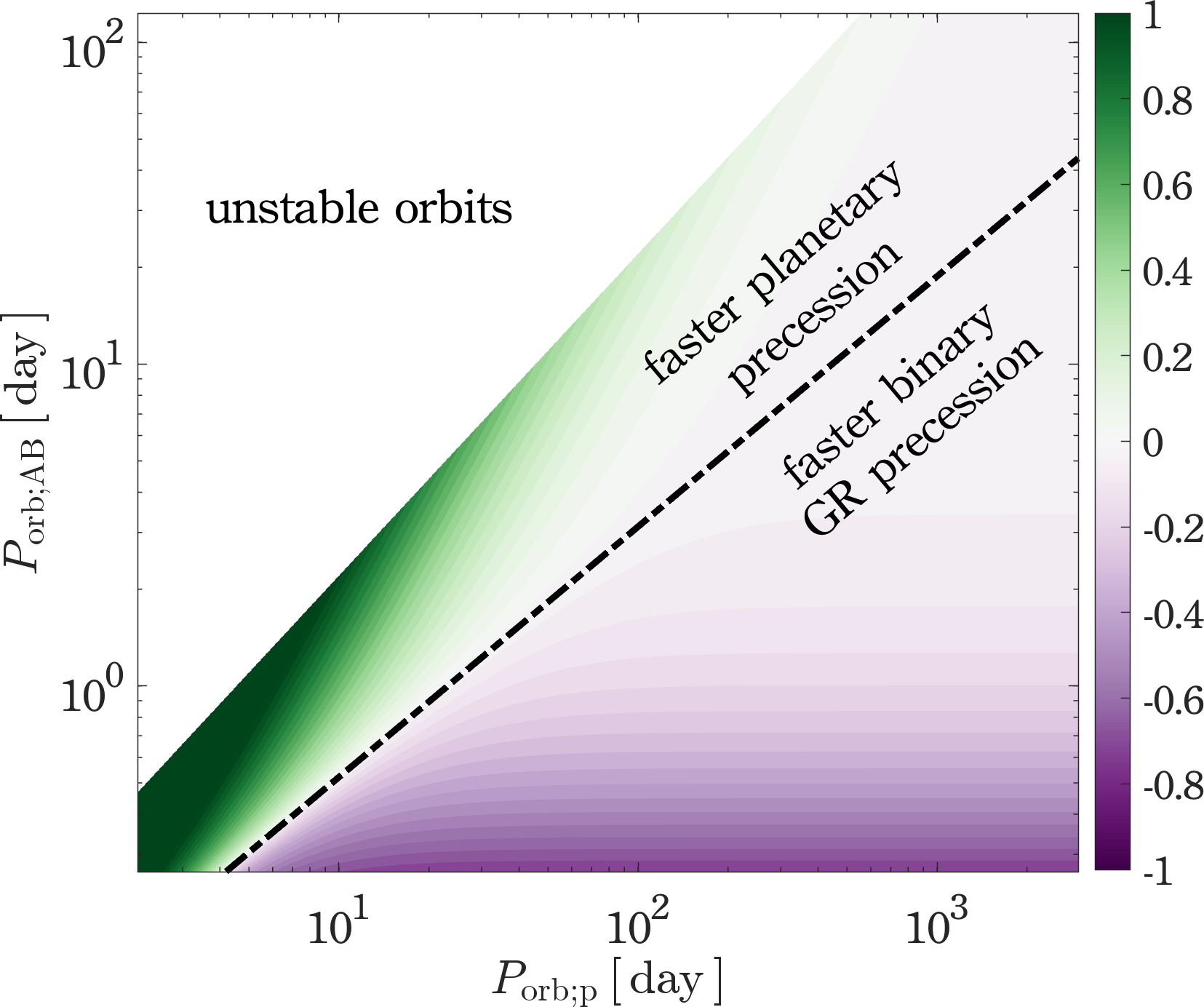}
\caption{Locating the resonance in orbital periods parameter space. Plotted is a contour surface of the function $\text{sgn}(\dot{\varpi}_{\rm p}-\dot{\varpi}_{\rm AB})\times\log_{10}(1+|\dot{\varpi}_{\rm p}-\dot{\varpi}_{\rm AB}|).$ The precession rates of the host binary and the CBP equate at the dashed black curve, identifying the resonance. We adopt the test particle approximation for the planet here, therefore $\dot{\varpi}_{\rm AB}$ is driven by GR, while $\dot{\varpi}_{\rm p}$ is driven by the binary's torque. We use $m_{\rm A} = 2M_{\odot}$, $m_{\rm B} = 1M_{\odot}$, and $e_{\rm AB} = 0.2$. The instability zone on the upper left follows the criterion of \cite{holman1999long}.    }

\label{Fig_contour_frequencies}
\end{figure}

Figure \ref{Fig_contour_frequencies} shows the difference between binary and planet apse precession rates, for a fiducial binary system, in the space of orbital periods of the binary and the planet. The condition of \eq{Resonance_condition} is satisfied on the dashed black curve, above which the binary forces the planet to precess faster than itself, and below which GR maintains a faster binary precession. Crossing from one region to the other guarantees encounter with the resonance. Whether in the crossing, planet and binary get caught in the resonance, i.e. continue evolving while locked into an apse synchronized configuration, 
is as it turns out a probabilistic matter, the study of which  takes us into the rich history of emergence of ubiquitous resonances in Solar system dynamics \citep[][]{gs1965,yoder1979diagrammatic,henrard1982capture,henrard1983second,borderies1984simple,malhotra1990capture}]. The upshot of it all is that capture or not is surely dependent on: the direction of the crossing, how \textit{slow or fast} it is, and the location of the system in phase space as it prepares to cross the resonance, with analytic expressions for probabilities of capture provided in the limit where system parameters evolve adiabatically, i.e. over timescales which are very slow when compared to timescales which are germane to the resonance in question.

So what are the features of the GR-driven apse resonance which concern us here, and how do they evolve with non-conservative processes of interest?

We are mainly interested in CBP dynamics around a tidally evolving binary, an evolutionary scenario which favors encounter then capture into extended periods of resonant forcing. Tidal evolution of the binary brings about changes in the orbital parameters of the binary which render the Hamiltonian governing the coupling between binary and planet time-varying, non-autonomous. In regimes where tidal evolution of the binary occurs over timescales significantly longer than any dynamical timescale in the three body dance (and this is certainly the case here), much of what is to be expected in the encounter of the resonance can be gleaned by examining the structure of the time frozen Hamiltonian over a selection of system parameters. 

The orbit-averaged dynamics of a test-particle which is coplanar with a relativistic, uniformly precessing, binary reduces to a one-degree of freedom Hamiltonian whose phase space structure is rather trivial to recover. We will map the evolution of the resonance, and with it the stable relative equilibria in which the test particle planet is synchronously locked in precession with the binary.

\begin{figure*}[ht]
\includegraphics[width=\textwidth]{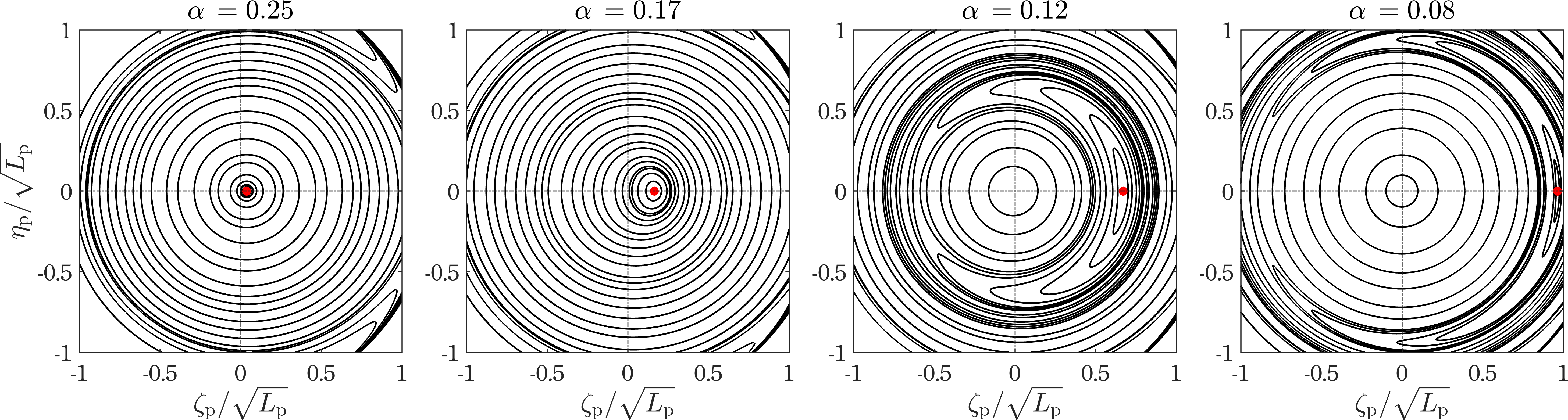}
\caption{Phase portraits in $(\zeta_{\rm p},\eta_{\rm p})-$space corresponding to the Hamiltonian of \eq{Hamiltonian_1}, which governs the planar dynamics of a circumbinary test particle planet around a stellar binary precessing under GR. The phase portraits are restricted to the region around the resonance of interest. Moving from left to right, level curves of the Hamiltonian are plotted in different panels for decreasing values of the drift parameter $\alpha=a_{\rm AB}/a_{\rm p}$, i.e., showing the phase space of the CBP as the binary's orbit shrinks. Beyond a critical value of $\alpha$, the secular precession resonance understudy is encountered, giving rise to the libration island around apse alignment  ($\varpi_{\rm p}=\varpi_{\rm AB}; \eta_{\rm p}=0)$, with the latter growing in eccentricity as $\alpha$ decreases further. At the center of the libration island lies the equilibrium point, shown in red, at which  $\dot{\varpi}_{\rm p} = \dot{\varpi}_{\rm AB}$, located by solving the equilibrium condition of \eq{equilibria_de_planar}. Here we adopted $m_{\rm A} = 2M_{\odot}$, $m_{\rm B} = 1M_{\odot}$, and $e_{\rm AB} = 0.35$.}
\label{Figure_CBP_phase_space}
\end{figure*}

We parameterize the motion of the CBP  by the normalized angular momentum vector $\bold{j}_{\rm p}=\sqrt{1-e_{\rm p}^2}\hat{\bold{n}}_{\rm p}$, and the Lenz vector $\bold{e}_{\rm p}=e_{\rm p}\hat{\bold{u}}_{\rm p}$, where $e_{\rm p}$ is the CBP's eccentricity, while $\hat{\bold{n}}_{\rm p}$ and $\hat{\bold{u}}_{\rm p}$ denote the directions of the CBP's orbital angular momentum and periapse respectively. Analogous definitions are adopted for the binary. Using these vectorial elements, it is straightforward to recover the Hamiltonian governing the 
orbit averaged dynamics of the CBP,  generalizing related works on hierarchical triples \citep[see Appendix; e.g.,][]{tremaine2009satellite,correia2011tidal,hamers2020secular,farhat2021laplace,farhat2023case}. Under the test particle approximation, the binary’s GR induced precession is uniform \footnote{We neglect loss of binary energy and angular momentum with radiation of gravitational waves, a process which is negligibly slow for the orbital periods under consideration.}, suggesting transformation to a frame co-rotating with the binary, ending up with the following amended Hamiltonian for the CBP’s dynamics
\begin{equation}\label{Hamiltonian_1}
       H_{\rm S}=\frac{Gm_{\rm AB}}{a_{\rm p}}\Psi_{\rm N}-\dot{\varpi}_{\rm AB} \sqrt{G m_{\rm AB}a_{\rm p} }\,( \bold{j}_{\rm p}\cdot\bold{\hat n}_{\rm AB}).
 \end{equation}
Constant Keplerian terms were dropped; the Newtonian gravitational interaction between the inner binary and the CBP is captured to octupolar order through the averaged disturbing function $\Psi_{\rm N}$, i.e., $\Psi_{\rm N}=\left(\Psi_{\rm N,quad}+\Psi_{\rm N,oct}\right)$; the precession rate of the binary is given by
\begin{equation}
    \dot{\varpi}_{\rm AB} = \dot{\varpi}_{\rm GR} = \frac{3Gm_{\rm AB}}{c^2 a_{\rm AB}} \frac{n_{\rm AB}}{1-e_{\rm AB}^2},
\end{equation}
with $n_{\rm AB}$ being the mean motion of the binary and $c$ the speed of light.  

Figure \ref{Figure_CBP_phase_space} renders coplanar phase-space dynamics as governed by $H_{\rm S}$ using canonical coordinates $(\zeta_{\rm p},\eta_{\rm p})=\sqrt{2L_{\rm p}\left(1-\sqrt{{1-e_{\rm p}^2}}\right)}\left(\cos\Delta\varpi,\sin\Delta\varpi\right)$ over a range of $\alpha=a_{\rm AB}/a_{\rm p}$. We highlight the feature which is critical to our process, an island of stable, resonant, libration, centered at $\eta_{\rm p}=0$,  i.e. around apse-aligned ($\Delta\varpi=0^\circ$) resonantly locked configuration. The center of stable libration grows in eccentricity with decreasing $\alpha$: indeed with a shrinking binary, GR induced precession is faster, and a planet will need to be on a sufficiently eccentric orbit for its forced precession to match the binary’s. The width of the island, which is nonlinearly sculpted by a separatrix, appears to increase away from inception of the resonance; then with the shrinking binary, it decreases as the island approaches a high eccentricity counter-part, with a conspicuous unstable aligned configuration. Beyond these resonant domains, the planet’s eccentricity evolves over an integrable phase-space. 

The aligned resonantly locked configurations, which are marked in red in Figure \ref{Figure_CBP_phase_space}, are stationary solutions of the dynamics generated by $H_{\rm S}$ which we recover by setting: $d\bold{j}_{\rm p}/dt = d\bold{e}_{\rm p}/dt =0$, then further selecting  $\Delta\varpi=0^\circ$. Restricting to coplanar configurations, we end up with:
\begin{align}
\nonumber
    0&= \frac{9\varepsilon_{\rm q}}{8}\frac{ e_{\rm p}}{(1-e_{\rm p}^2)^2}\left[e_{\rm AB}^2+2/3\right]\\ \nonumber
    &-\frac{15\varepsilon_{\rm q} \varepsilon_{\rm o}}{64}\frac{(4e_{\rm p}^2+1)}{(1-e_{\rm p}^2)^3}e_{\rm AB}  \left(3e_{\rm AB}^2+4 \right)\cos\Delta\varpi\\
    & -\dot\varpi_{\rm GR }e_{\rm p} \cos \Delta\varpi.\label{equilibria_de_planar}
\end{align}
Here, $\varepsilon_{\rm q}$ and $\varepsilon_{\rm o}$ are the dimensionless coefficients characterizing the strength of the quadrupolar and octupolar components of the expansion of the potential $\Psi_{\rm N}$ (Appendix,  Eqs. \ref{Psi_N_quad}-\ref{Psi_N_oct}), and $\Delta\varpi = \varpi_{\rm AB} - \varpi_{\rm p}.$ 

\begin{figure}[t]
\includegraphics[width=.45\textwidth]{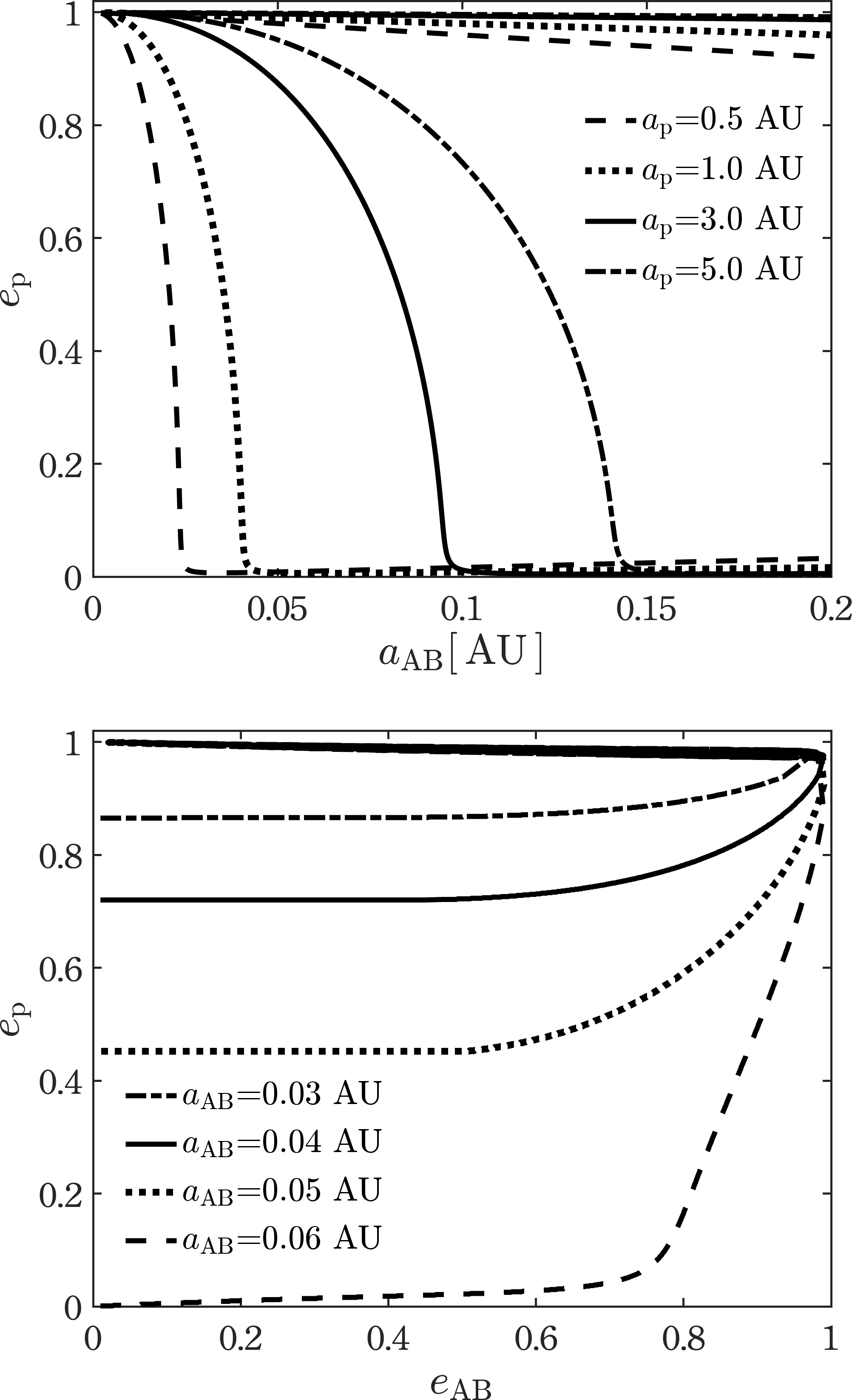}
\caption{ Profiles of CBP equilibrium eccentricity when apse-aligned with the binary: (Top) at different locations for the binary and the planet with $e_{\rm AB}=0.2$; (Bottom) for different binary location and eccentricity with $a_{\rm p}=1.5$~AU.  The realistic dynamical behavior of a CBP is expected to develop around this equilibrium structure if the system is captured in resonance and evolves in time while maintaining adiabaticity. Here we have adopted $m_{\rm A} = 2M_{\odot}$, $m_{\rm B} = 1M_{\odot}$, while the CBP is treated as a test particle.}
\label{Fig_equilibria}
\end{figure}

Over a suitable range of $a_{\rm AB}$, $e_{\rm AB}$ and for a given $a_{\rm p}$, the solution of \eq{equilibria_de_planar} yields families of equilibria, labeled by the equilibrium $e_{\rm p}$, a sample of which is rendered in Figure 
\ref{Fig_equilibria}. In the top panel, for a fixed $e_{\rm AB}$ and  $a_{\rm p}$, and with varying binary semi-major axes, a family of stable equilibria bifurcates from circular orbits as the resonance is first encountered, becoming nearly radial as it meets the the high eccentricity branch of unstable aligned equilibria.  The lower panel considers a complementary regime, whereby the binary circularizes at fixed $a_{\rm AB}$ (with all planets having the same $a_{\rm p}=1.5$ AU). Equilibria bifurcate at large $e_{\rm AB}$ and  $e_{\rm p}$ into two families, one unstable which follows the high eccentricity branch, and the other stable decreasing almost linearly with decaying binary eccentricity before plateauing at a value which decreases with increasing $a_{\rm AB}$. 

The high eccentricity branches shall not concern us in what follows: they obtain at eccentricities where the multipole expansions which we are working with are suspect (in terms of accuracy and convergence), and where planetary orbits are sure to get disrupted by the binary over faster MMR related timescales. 

Focusing on the stable low eccentricity branch, one can go beyond equilibria to consider dynamics around the island. Of prime importance to what follows are:
\begin{itemize}
    \item The timescale of small amplitude libration around the equilibrium, as the island migrates with binary shrinkage. Here we note that for the same fiducial system with a test-particle planet around 1.5~AU the timescale in question evolves non-monotonically in the range $[10^3, ~1.6\times10^5]$ yrs, with the maximum reached around an $\alpha= \frac{a_{AB}}{a_p} = 0.13$, before decreasing again with the shrinking binary to around $4\times10^3$ yrs around $\alpha=0.03$.
    \item Of equal and perhaps more subtle importance is the evolution of the islands size as measured through the width of its separatrix. The evolution of the separatrix width (and the enclosed area along with it) shape both the probability of capture, as well as transport of a distribution of planets from some initial formation determined state to one sculpted by passage of, capture in, excitation by, then release from resonance  \citep[refer to][for details on adiabatic sculpting of distributions]{st1996}. For the process under consideration, width evolution is non-monotonic, reaching its maximum in sync with the libration timescale, then decaying with the shrinking binary, and with island center eventually at near radial position.
\end{itemize}

Taken together, these skeletal structures (idealized though they may be) shape the dynamics of a planet which encounters this seemingly innocuous resonance in the course of non-conservative evolution of the binary’s orbit, whether through tidal evolution as we consider below or through other processes.

Theory tells us that capture into resonance is certain provided: (\textit i) resonance sweeping occurs on time scales slow enough for the traversal of the resonance width in phase space to take longer than the typical libration period of the planet in resonance (longer than $10^5$ yrs for the case we considered, with an order of magnitude above or below for more distant or close-in planets); (\textit{ii}) the resonant island drifts in the direction of increasing eccentricity (typically resulting from decay of binary semi-major axis, or outward migration of the planet or both); and (\textit{iii}) the area enclosed by the planet’s trajectory in phase space is smaller than the area enclosed by the separatrix which, in the case of interest to us here, translates to a condition on $e_{\rm p}$ being initially smaller than a critical value.

Under ideal circumstances,  we are sure that a CBP starting with a near-circular orbit, around a binary whose semi major axis is decreasing over timescales significantly longer than the resonant libration timescale, will get caught in the GR induced resonance as it crosses it, and find its eccentricity pumped to nearly radial orbits as it tracks the corresponding family in  Figure 
\ref{Fig_equilibria} [upper panel]. Thus it seems that sustained capture with a binary on its course to merger is destined to get rid of the planet, either by engulfing it into the binary, or by destabilizing  and ejecting it (hence our excitement about the process!). This is so with binary semi-major axis decreasing adiabatically at constant binary eccentricity. Should the process (and this is unlikely) allow eccentricity decay at constant semi-major axis, capture will surely be unlikely with the island migrating in the opposite direction for capture to occur, whether adiabatic or not. The more likely scenario is for eccentricity damping to accompany semi-major axis decay, and here things can get complicated with CBP eccentricity forcing with binary orbit shrinkage countered by eccentricity decrease with binary eccentricity damping.

Our analysis of the idealized behavior will prove handy in qualifying (if not outright interpreting) the outcomes of the more complex simulations we report below.

\section{Caught in GR-induced Resonance: CBP around tidally evolving Binary} 
\label{Section_tidal_evolution}

We animate the quasi-static structures delineated above by considering a CBP's response to a GR-induced apsidal precession resonance as it sweeps past its orbit in the course of an in-spiralling binary.  Orbital decay in compact binaries ($P_{\rm orb;AB}\,{\sim}10^0{-}10^2$~days) can be driven by tidal dissipation within the binary, magnetic braking, GR-induced gravitational waves emission, or a combination thereof. However, these mechanisms are effective over different spatial- and time-scales:  GR-induced decay is effectively the slowest \citep[the decay timescale of solar-like binary components on a 7-day orbit is ${\sim}10^4~{\rm Gyr}$; e.g., ][]{peters1964gravitational}, while magnetic braking requires the orbital period of solar-like stars to be less than 5 days for the  timescale to be comparable to Hubble time \citep[e.g.,][]{krishnamurthi1997theoretical,Chaboyer1995,el2022magnetic}. 

Here, and while keen on exploring the implications of both processes, we instead favor a natural, fairly efficient, and robust mechanism which relies on orbital decay driven by tides, and is effective over a --relatively-- wider range of orbital separations and timescales. Stellar tides act on timescales characterized by
\begin{equation}
       \tau_{\rm tide} = \frac{m_{\rm A}m_{\rm B}}{m_{\rm A}+m_{\rm B}}\frac{n_{\rm AB}^2a_{\rm AB}^2}{2\dot{E}_{\rm diss}},
\end{equation}
with $\dot{E}_{\rm diss}$ being the rate of tidal energy dissipation. The latter derives from a combination of: the equilibrium tidal response in the stars, which dissipates energy via turbulent eddy viscosity of tidally driven shear in the stellar convective envelope \citep[][]{zahn1977tidal,zahn1989tidal}; and the dynamical tidal response, a series of tidal modes, mainly gravity waves, excited at the radiative-convective boundary, which can efficiently dissipate energy when damped, for instance, by radiative diffusion \citep[e.g.,][]{savonije1983tidal,goodman1998dynamical,sun2018orbital}. These oscillatory modes can also be resonantly excited, amplifying the rate of energy dissipation, and consequently the rate of orbital decay \citep[e.g.,][]{terquem1998tidal,fuller2016resonance,zanazzi2021tidal}. As such, the timescale of tidal orbital decay can vary over several orders of magnitude depending on stellar interior, spin-alignment, and orbital eccentricity, allowing for the formation of tight binaries with $P_{\rm orb;AB}\,{\leq}7$~days from wider orbits within $10^8{-}10^{10}$~yrs. With the resonant libration timescale being $\leq10^5$~yrs, this setting offers a rich space for exploring  the fate of CBPs encountering the resonance understudy under favorable, adiabatic conditions. 

\subsection{Planets around tidally decaying binaries: Population synthesis}
To get a sense of how the secular resonance can sculpt a typical distribution of CBPs, we perform a suite of simulations tracing the  dynamical evolution of triple systems, each including a stellar binary and a CBP, while allowing for: \textit{i)} the mutual Newtonian perturbative interaction between the binary and the planet, captured to octupolar order,  \textit{ii)} the first order post-Newtonian gravitational effect of the binary, \textit{iii)} the tidal interaction between the binary components, driven by eccentricity tides, and driving in return orbital circularization and decay, and \textit{iv)} the tidal interaction between the binary and the planet. The secular equations of motion used to model these effects are delineated in the  Appendix.

{We select our binary systems from the Villanova Kepler Eclipsing Binary Catalogue\footnote{https://keplerebs.villanova.edu}\citep[][]{prvsa2011kepler,kirk2016kepler}, and exclude binaries with strongly distorted light curves by requiring a morphology parameter \textsc{Morph} $<0.6$. Low \textsc{Morph} values typically identify well-detached systems with light curves showing clearly separated eclipses, whereas higher values correspond to overcontact or ellipsoidal binaries \citep[][]{matijevivc2012kepler}. We focus on the former, presuming that their present separations were set primarily by tidal decay from initially wider orbits. Overcontact systems likely reflect additional evolutionary drivers such as angular momentum loss under magnetic braking \citep[e.g.,][]{verbunt1981magnetic,gazeas2008angular}, as we discuss in Section \ref{Section_Conclusion}.}

{The detached population contains 1630 systems with known present periods. To infer their initial periods, we assume that tidal decay and circularization proceeded at constant orbital angular momentum: for each binary, we solve for an initially wider and eccentric progenitor which would evolve tidally to the observed configuration. The exercise requires us to specify the eccentricity of that progenitor.} Recent studies on the interaction between accreting binaries and their surrounding discs reported the existence of equilibrium attractors in the binary's eccentricity at $e_{\rm AB}=0$, for expanding orbits, and $e_{\rm AB}{\sim 0.4{-}0.5}$ for shrinking orbits \citep[][]{zrake2021equilibrium,dOrazio2021orbital,siwek2023preferential,valli2024long}. As such, our binaries' initial eccentricities are assigned from a Beta distribution with $\alpha=1.75,\beta=2.01$ \citep[][]{price2020close}, with a mean at $e_{\rm AB,0}=0.47$.

The initial binary separation thus obtained is not to be misconstrued as the separation at formation by fragmentation of circumstellar discs or protostellar cores. Indeed, it is generally argued that most binaries form on wider ($10{-}1000~{\rm AU})$ orbits and migrate inwards, driven by gas dynamical friction, torques arising from the disc, protostellar accretion, interactions with a tertiary star on a wider orbit, or a combination thereof \citep[e.g.,][]{artymowicz1991effect,bate1995modelling,bate1997accretion,bate2002formation,eggleton2006mechanism,moe2018dynamical,tokovinin2020formation,valli2024long,offner2022origin}. Our initial distribution can therefore be taken as the outcome of formation and subsequent fast inward migration, leading to separations where tidal evolution takes over. 

The age of each of these systems is drawn from a uniform distribution between 1 and 10 Gyrs, and the mass of the primary is set to $1M_\odot$, while the secondary to primary mass ratio $q$ is drawn from a linear distribution $\propto q$, abiding by the widely reported excess of stellar twins on compact orbits \citep[e.g.,][]{tokovinin2000origin,elbadry2019discovery}. Stellar radii are calculated as a function of the masses using the scaling relations obtained by \citet[][]{choi2016mesa}. As for the planets, each is characterized with a mass $m_{\rm p}=1M_{\rm J}$, $e_{\rm p;0}=10^{-5}$, a mutual inclination with the binary's plane that is drawn from a uniform distribution over the interval [$0^\circ$, $3^\circ$], and a semi-major axis that is drawn from a uniform distribution over the interval [0.5, 3] AU. However, we condition our sampling such that each planet starts outside the instability zone of the associated binary. Hereafter, we use the empirical formula obtained by \citet[][their Equation 10]{georgakarakos2024empirical} for the critical distance defining the circumbinary instability zone, since it takes into account the effect of planetary eccentricity. 

With masses, age and initial conditions specified, we integrate the equations of motion (see Appendix) governing the secular, orbit averaged, evolution of each of these systems. As we are more interested in the dynamical fate of the CBP than that of the stars, the integration is stopped when any of the following conditions is met: the planet is ejected from the system; the binary circularizes and is no longer decaying in orbit; or the evolution reaches the prescribed age of the system.

\subsection{From synthesis to disruption: A first look}
\begin{figure}[t]
    \includegraphics[width=.47\textwidth]{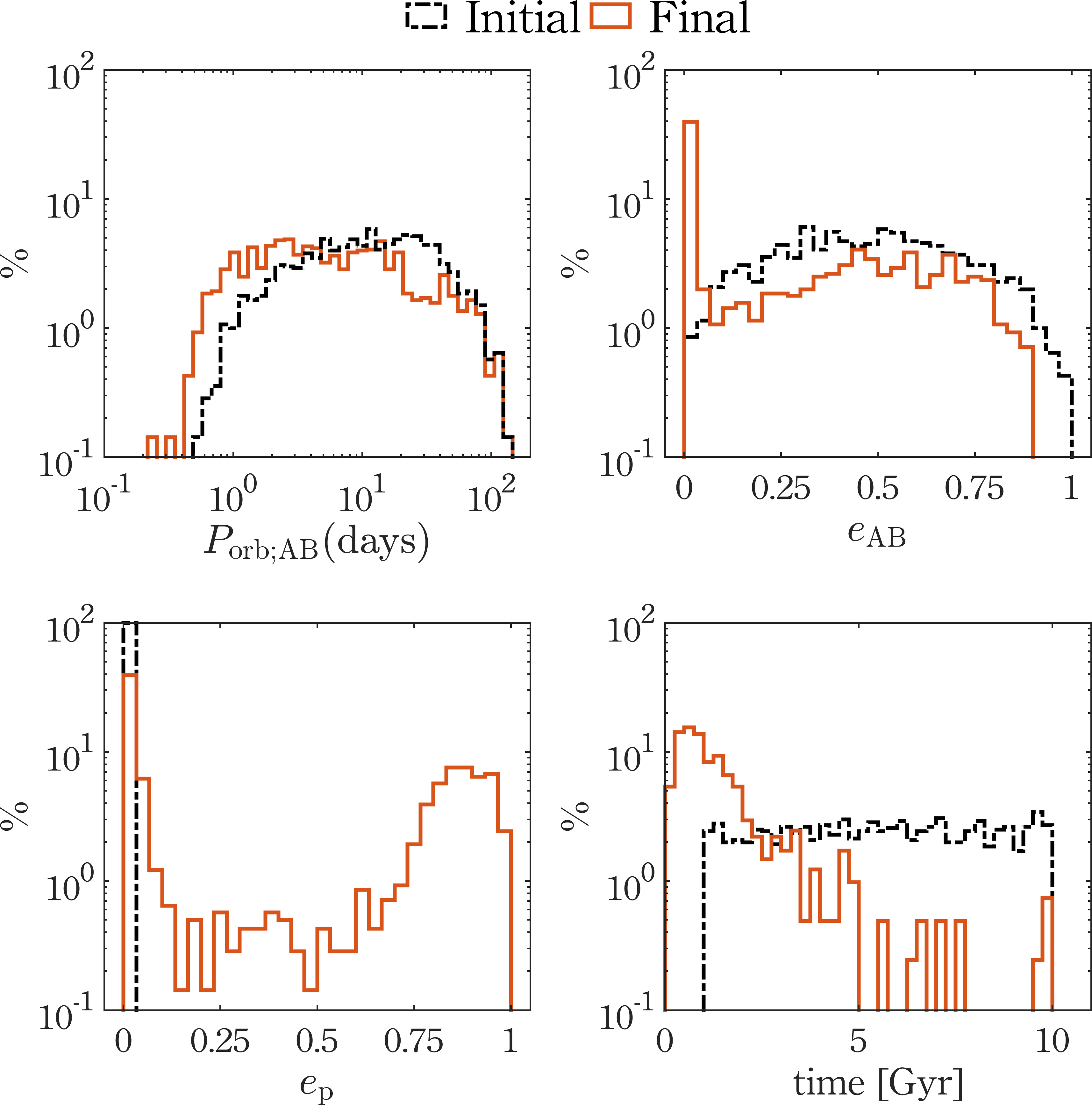}
\caption{Distributions of the initial and end states of the dynamical evolution of 1630 triple systems of binaries and CBPs. The population synthesis is described in the main text and the secular evolution follows the equations of motion in Appendix \ref{Appendix_EQM}. Shown, in order, are the population's initial and final binary orbital period $P_{\rm orb;AB}$, binary eccentricity $e_{\rm AB}$, planetary eccentricity $e_{\rm p}$, and the systems prescribed ages compared to the times at which unstable planets are destabilized. The binaries orbital decay and eccentricity decrease are driven by the binary tides, while the planets' eccentricity growth and probable destabilization are driven by capture into the secular apsidal precession resonance.}
\label{Fig_Final_tidal_distributions}
\end{figure}

In Figure \ref{Fig_Final_tidal_distributions}, we present the distributions describing the initial and end states of our simulations. {With the majority of binaries starting with considerably eccentric orbits, strong eccentricity tides drive efficient orbital decay, leaving behind  ${\sim}53\%$ of the population with $P_{\rm orb; AB}\leq 7.45~{\rm days}$\footnote{Hereafter, we refer to binaries with $P_{\rm orb; AB}\leq 7.45~{\rm days}$, binaries that observationally lack transiting planets, as tight binaries.} by the end of the simulations. As shown in the second panel of the figure, nearly $40\%$ of these binaries completely circularize their orbits. The rest find their orbital eccentricities decaying, with the population's mean eccentricity shifting from $\langle e_{\rm AB; 0}\rangle=0.47$ to $\langle e_{\rm AB; f}\rangle=0.24$.} 

\begin{figure*}[t]
\includegraphics[width=\textwidth]{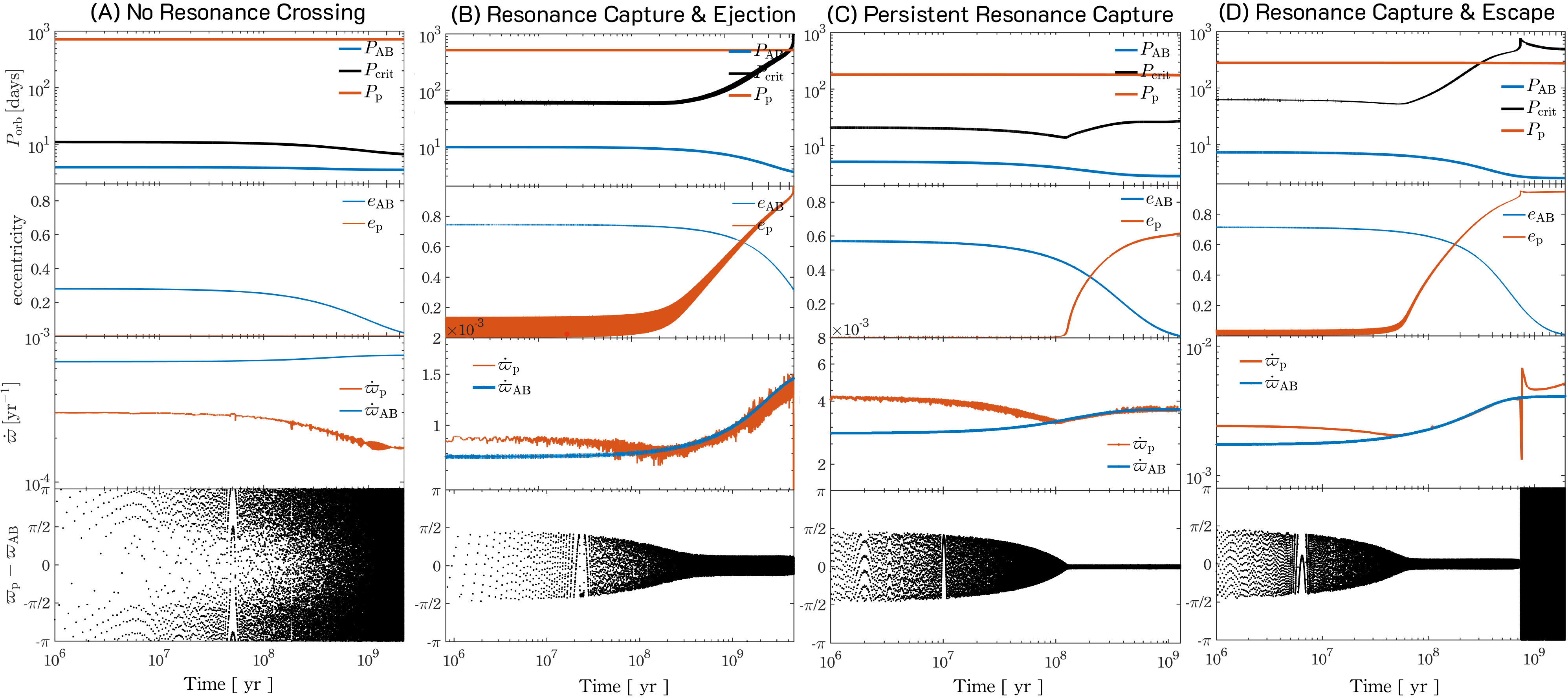}\caption{Sample simulations representing the outcome regimes of the population synthesis study of Figure \ref{Fig_Final_tidal_distributions}. For each regime, the first panel plots the evolution of the orbital periods of the binary, the planet, and the boundary of the circumbinary instability zone \citep[estimated following][]{georgakarakos2024empirical}; the second plots the eccentricity evolution of the binary and the planet; the third plots the evolution of the rate of change of the longitude of the periaspse for both the binary and the planet; while the fourth plots the resonant angle $\varpi_{\rm p}-\varpi_{\rm AB}$. {From left to right, $m_{\rm AB}=2.28, 1.58, 2.33,$ and $1.95\,M_{\odot}$. The first regime corresponds to CBPs that do not encounter the sweeping resonance in the course of the binary's tidal decay. Regime B showcases the resonant capture and subsequent ejection of the planet from the system. The third column shows resonance capture that leaves the planet on a stable eccentric orbit, saved by the binary stopping its orbital decay upon circularization. The fourth column shows the richer dynamics of Regime D where the planet is first captured by, and later escapes the resonance.  We elaborate further on the dynamical pathways of these regimes in the main text.  }}
\label{Fig_individual_sims_regimes}
\end{figure*}

{The third panel of Figure \ref{Fig_Final_tidal_distributions} captures the evolved distribution of CBPs. With all the planets initiated on circular orbits, our results predict that only ${\sim}38\%$ of the planets maintain their near circular orbits by the end of the simulation.} This family corresponds to a regime where the planet never encounters the sweeping secular resonance in the course of the binary's orbital decay. In this regime, which we label Regime (A), either the binary started its orbital decay with its orbital radius smaller than the ``resonant" orbit radius, or it started further out but circularized its orbit (or reached the system's age) before reaching that orbital radius, i.e. before the planet encountered the resonance. {We show an example of this regime in the first panel of Figure~\ref{Fig_individual_sims_regimes}. Therein, $P_{\rm orb,p}=720~{\rm days}$, and the binary's orbit shrinks starting from $P_{\rm orb,AB,0}\simeq11~{\rm days}$, i.e., inside the resonant orbital period that we estimate for this system's parameters to be ${\sim}17~{\rm days}.$} In this regime, the system starts with $\dot{\varpi}_{\rm AB}>\dot{\varpi}_{\rm p}$, with the two frequencies further diverging as the binary tightens. Consequently, the resonance is never crossed, the angle ($\varpi_{\rm p}-\varpi_{\rm AB})$ circulates throughout the system's lifetime, and the binary circularizes its orbit by $2.21~{\rm Gyr}$, leaving behind the planet on a stable circular orbit. 

{In contrast, ${\sim 55}\%$ of the full CBP population do encounter the resonance adiabatically as the binary slowly shrinks its orbit (the remaining 7\% find themselves with $e_{\rm p;0}<e_{\rm p}<0.1$ due to the non-resonant forcing of the binary).}  Before a system is caught in the resonance, orbital angular momentum shuttles back and forth between the binary and the planet, inducing limited, small amplitude oscillations in the eccentricity of each. Upon resonance capture, the reciprocal exchange of angular momentum is converted into a unidirectional flow from the planet to the binary, which steadily increases the planetary eccentricity. {More specifically, the outcome of this resonant eccentricity pumping in our simulations leaves ${\sim 14}\%$ of the planets on moderately eccentric orbits ($0.1\leq e_{\rm p}\leq 0.8)$, while the remaining resonant majority (${\sim 41}\%$) are pumped to extremely eccentric orbits with $ e_{\rm p}\geq 0.8.$ Out of the full population, ${\sim 39}\%$ (i.e. $71\%$ of the planets that do encounter the sweeping resonance) are forced, by virtue of their increased eccentricity, to protrude into the circumbinary instability zone at some point in the system's lifetime.}

In the last panel of Figure \ref{Fig_Final_tidal_distributions}, we restrict the population to the family of eccentric, destabilized planets, and we record the time at which these planets enter the instability zone. {With the black histogram showing the near-uniform distribution of prescribed ages for the systems, the orange histogram shows that ${\sim 53}\%$ of destabilized planets enter the instability zone within the first ${\rm Gyr}$.} Early onset of resonant capture (then destabilization) is critical to the chance of detecting transits, the latter being possible only after the binaries reach their photometric stability.

\subsection{Inside the resonance trap:  Spiral, Stir, or Survive}

For the population of planets that do encounter  the sweeping resonance, their dynamical fates can be broadly classified by three evolutionary regimes (B, C, and D), with a representative example of each plotted in Figure \ref{Fig_individual_sims_regimes}.

\begin{itemize}
    \item[{I})] In Regime B (second panel of Figure \ref{Fig_individual_sims_regimes}), the system starts with $\dot{\varpi}_{\rm AB}<\dot{\varpi}_{\rm p}$, then as the binary's orbit shrinks under tides, its GR-induced apsidal precession becomes faster, while its effect on the planet is weakened, slowing down the planet's precession. {For the system understudy, this leads to a match between the two frequencies within ${\sim}200{-}300~{\rm Myrs}$, allowing for capture into the secular resonance.} The system continues evolving with precession frequencies synchronized, as the binary's orbit continues to shrink and circularize. This requires simultaneously depleting the planet's angular momentum by continuously pumping its eccentricity until it attains a radial orbit, eventually ejected or engulfed by the binary. Capture is evident in the time evolution of the resonant angle, $\varpi_{\rm p}-\varpi_{\rm AB}$, with oscillation now restricted from  $[-\pi/2,\pi/2]$ to small amplitude librations around 0.

\item[{II})] {Regime C  (third panel of Figure \ref{Fig_individual_sims_regimes}) typically refers to planets that are captured into resonance as the binary decays, but end up with finite moderate-to-high eccentricity, rather than a (near-)radial orbit. These planets exhibit a behavior of resonant eccentricity growth  similar to that seen in Regime B, but with one key difference: the binary circularizes and stops its decay (or reaches its prescribed age) before the planet attains a radial orbit. The fate of the planet then hinges on whether its resonant eccentricity is sufficient for it to protrude into the instability zone before the system stops evolving. Indeed, and as shown in the top figure of Regime C, the zone starts shrinking with the binary's decaying and circularizing orbit, but then expands upon resonance capture, by virtue of the increasing planetary eccentricity. In the case understudy, the planet always remains outside of the instability zone, but this is not the case for the rest of CBPs in this regime.}

\item[{III})]{Finally, we identify a class of resonant systems  for which capture appears not to be persistent, but rather followed by resonance escape. Referred to as Regime D in Figure \ref{Fig_individual_sims_regimes}, the system initially evolves pretty much like in Regimes B \& C, with $\dot{\varpi}_{\rm p}$ approaching $\dot{\varpi}_{\rm AB}$ from above until the two frequencies synchronize and $e_{\rm p}$ starts growing. However, in contrast to Regimes B \& C where the planet remains captured with $e_{\rm p}$ increasing monotonically till the end of the simulation, here the planet appears to escape resonance with a large, now steady, $e_{\rm p}$.}

{The onset of this change of behavior can be understood in reference to the equilibrium structure shown in Figure \ref{Fig_equilibria}. Therein, we have shown that, while the planet’s equilibrium resonant eccentricity increases as  $a_{\rm AB}$ decreases, it also decreases with the decay of $e_{\rm AB}$. In Regime D, the circularization of the binary forces $e_{\rm p}$ to transition from an $\dot{a}_{\rm AB}$-dominated regime (where $e_{\rm p}$ tends to grow with ${a}_{\rm AB}$ decreasing) to an  $\dot{e}_{\rm AB}$-dominated regime (where $e_{\rm p}$ decreases with ${e}_{\rm AB}$ decreasing). This behavior accompanies the phase space evolution shown in Figure~\ref{Figure_CBP_phase_space}: while the resonant island grows in $e_{\rm p}$ as $a_{\rm AB}$ decreases, it also shrinks in size, allowing the planet to escape its libration zone at high $e_{\rm p}$ values.  As such,  the planet avoids being indefinitely pumped in eccentricity by escaping the resonance, and the resonant angle returns back to full range circulation, leaving behind the planet with a finite eccentricity. }

{While avoiding the fate of Regime B, the captured planet is evidently pushed into the binary's instability zone (Regime D, top panel). That zone initially shrinks with the shrinking and circularizing binary orbit, then rapidly expands upon resonance encounter and resulting growth of $e_{\rm p}$ to engulf the planetary orbit, then recedes slightly with the binary’s decreasing eccentricity and semi-major axis. Regime D is typical of a family of planets in our simulations which end up with moderate to high $e_{\rm p}$. The fate of these planets hinges on whether or not they encounter and survive their residence in a waxing and waning instability zone by simulation stopping point. In the case of the system featured in the fourth column of Figure~\ref{Fig_individual_sims_regimes}, the phase in question is hundreds of millions of years long, giving the planet plenty of time to stir out of the system. Removed for sure!}

\end{itemize}

\subsection{Spared yet hard to detect}
\begin{figure}[t]
\includegraphics[width=.47\textwidth]{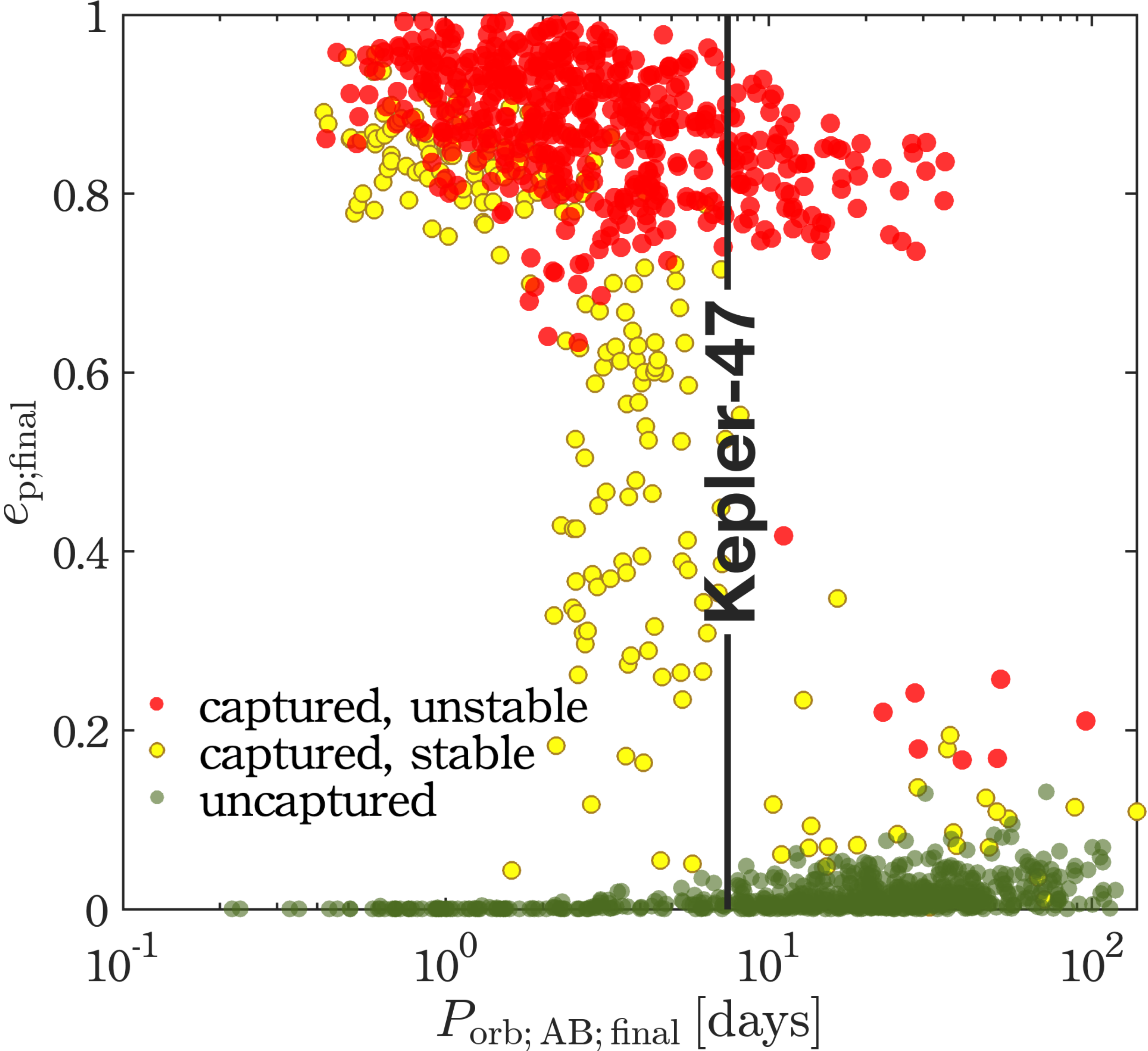}
\caption{Resonant decimation of circumbinary planets. From our population study of Figure \ref{Fig_Final_tidal_distributions}, we plot the planetary eccentricity as a function of the binary's orbital period at the end of our simulations. All CBPs were initiated on circular orbits. Three distinct families are identified in the plot: $\textit{i)}$ in green, a family of planets that avoid the sweeping resonance and thus maintain stable near-circular orbits; $\textit{ii)}$ in red, a family of planets captured in resonance and consequently excited to extreme eccentricity values, to be destabilized and ejected, as their host binaries reach tight orbits; and $\textit{iii)}$ in yellow, a family of planets which are resonantly excited but end up on stable eccentric orbits, saved from destabilization by the binary reaching its tidal equilibrium early enough, or by occupying wide orbits. The orbital period of the tightest binary observed to host a transiting planet, Kepler-47, is marked by the vertical line.
}
\label{Fig_final_eccentricities}
\end{figure}

We bring our simulations in closer contact with observables by correlating the orbital period of tidally evolved binaries with the ultimate eccentricity of their CBPs. The resulting Figure \ref{Fig_final_eccentricities} provides a vivid synthesis of survivability through our process,  then insights into the potential detectability of surviving planets. 

Marked in green is the family of planets which avoided the sweeping resonance in the course of the binary's decay, thus maintaining near-circular orbits, with slight eccentricity excitation due to secular forcing by their sloshing hosts. As evident in Figure \ref{Fig_final_eccentricities}, those uncaptured planets are spread over a broad range of binary periods with a notable concentration trend: planets revolving around (relatively) wide binaries are more likely than not to be spared by the resonance and its nefarious perturbations.

Then, we have the planets which were caught in the GR-driven resonance, and which fall into two distinct families: \textit{i}) red for those planets (around mainly tight binaries) which were resonantly driven to extreme eccentricities, thus penetrating the circumbinary instability zone and facing near certain disruption, whether through direct encounter with the binary or ejection from the system; and \textit{ii}) yellow for planets which were captured in resonance, experienced moderate to large eccentricity growth without entering the host's instability zone\footnote{
We remind the reader that simulations were stopped when either the binary is circularized, or the system reached the prescribed age, leaving the question of the long-term, non-secular stability and survival of remnant eccentric planets open to further analysis, both dynamical and astrophysical.}.

Around tight binaries, ones which are tidally driven to $P_{\rm orb;AB, final}\leq7.45~{\rm days}$ (that of Kepler-47), $82\%$ of the planets are resonantly excited (yellow and red), with $76\%$ entering the binary's instability zone (red). The ominously reddish mushroom in this critical period range provides as vivid a smoking gun as one can hope for in favor of our proposed pathway to the decimation (hence dearth) of CBPs around tight binaries. Naturally, and given the probabilistic nature of the process, from initial conditions to resonance sweeping itself, we do not expect our proposition to be foolproof: the yellow stem (with $0.1\lesssim e_{\rm p}\lesssim0.8$) together with the green scatter of Fig.~\ref{Fig_final_eccentricities} are here to remind us of that\footnote{In addition to the Kepler-motivated distribution of binary periods, we simulated a population where a binary's initial separation is sampled from a log-uniform distribution \citep[as in][]{korntreff2012towards,tokovinin2020formation,zanazzi2022tale} with $5 \leq P_{\rm orb; AB, 0}\leq 60~{\rm days}$. We verified that the proposed mechanism remains effective for this distribution, with capture into resonance destabilizing a comparable fraction of CBPs ($\sim 71\%$) around tight binaries [$P_{\rm orb; AB, final} \leq 7.45~{\rm days}$].}!

What about the detectability of surviving planets around those same tight binaries? Survivors fall into two distinct groups: green for planets which do not encounter the resonance in the course of binary in-spiral, and yellow for planets remote enough that resonant growth of eccentricity would still leave them outside the instability zone [this is to be contrasted with the negligible chance of planet survival in compact systems, since capture here is almost certain and resonant eccentricity growth will almost always drive captured planets into the instability zone]. Probing the semi-major axis distribution of survivors around tight binaries, we note that the odds of survival increase with widening orbits, reaching ${\sim}31\%$ by 3~AU, with $\langle a_{\rm p}\rangle = 2.43$~AU. What about their orbital inclination? Having initiated planets with $i_{\rm p}\in[0^\circ, 3^\circ]$, we find that they all maintained their near-coplanar relation to the host binary in the course of resonant coupling. Such nearly-coplanar, eccentric planets will certainly transit binary hosts in the Kepler catalog, but the likelihood of detecting their transit is rather low, it being inversely proportional to their orbital period which is rather long \citep[see discussion in Section \ref{Section_Conclusion}; e.g.,][]{welsh2012transiting,martin2015circumbinary,li2016uncovering}. In sum, planets which are most likely to survive the resonance (mainly distant ones) are least likely to have their transits detected, allowing us to safely conclude:  binaries which reached their tight orbits through tidal decay would almost surely clear the zone of detectable transiting planets by resonantly forcing the planets which inhabit that zone out of the system.

\section{Process and Precursors}
\epigraph{\textit{The fact is that each writer creates his precursors. His work modifies our conception of the past, as it will modify the future.}}{J. L. Borges, \textit{Kafka and his precursors} (1951)}
    
We have brought tidal evolution towards a tighter binary in dialogue with a secular resonance whereby a relativistic binary finds itself precessing at the frequency with which that binary forces precession in a companion planet. In the course of this dialogue, and as the binary decays under tidal migration, the planet is entrained by the binary's dance and finds its orbit stimulated to higher eccentricity as it remains in sync with an in-spiraling, faster-precessing binary. The orchestration of this dance to explain the occurrence of planets around tight binaries is to be sure thoroughly original to this work, however in its coming together, it forced a careful examination of (exo)-planetary precursors to our GR-induced apsidal resonance. And sure enough, precursors there where, and are here organized for our readers, in preparation for a future comprehensive treatment. 

In this context, we found it most helpful to begin with the closest manifestation in \citet[][]{mg2009}, revisit their references and related works, then highlight early indicators in Solar System dynamics: 
\begin{itemize}
    \item \citet[][]{mg2009} consider the effect of general relativistic precession together with quadrupolar contributions from a stellar bulge and/or close-in mass distribution on  the secular dynamics of a hierarchical planetary system. They isolate a GR+quadrupole induced secular resonance, which is precisely the one discussed here, exploring the associated reduced phase space and its bifurcations.
    
    \item \citet[][]{mard2007} and \citet[][]{ml2004} are concerned with the survival of close-in planets themselves subject to tidal evolution in the presence of GR precession. Their work builds on an earlier realization that GR precession can suppress Kozai cycling, and they are here exploring the same in coplanar configurations, with a two planet system passing through a secular resonance. Tidal evolution of the inner planet guides it towards a stable equilibrium balancing eccentricity tides, with resonant forcing by the external planet. They demonstrate how GR precession contributes to reducing that equilibrium eccentricity. Otherwise little mention of GR induced resonance per say, and the deformation it brings into the nonlinear phase space of the two planet system. 
    \item Digging deeper, yet another antecedent arises, this time in the triaxial Kozai setting, with \citet[][]{fkr2000} considering the effect of resonance between the general relativistic precession of a binary pulsar (PSR B1620 26 system) and secular, Newtonian, precession in the system. This is precisely the resonance which concerns us here only considered in spatial simulations, with minimal analysis of the processes at work. Their section 4.3 reports on the location and strength of the resonance, as a function of inclination of the planet, and its semi-major axis. We confirm in our analysis that a coplanar 2 Jupiter mass planet would have to be around 48 AU for exact GR driven apsidal resonance. We further explore implications of binary migration, and recover excitation of binary eccentricity during passage through resonance, with a corresponding decrease in planet eccentricity.
    \item This is as far as single planets with inner and/or outer perturbers are concerned. General relativistic effects in multiplanet systems have an equally rich trail, with mainly explicit references to the role of GR precession in tuning/de-tuning of secular resonances involving Laplace-Lagrange modes of the planetary system in question. The trail is a multilayered one, and largely rooted in Solar System dynamics, to which we restrict ourselves here. We start with what to our knowledge is the first implication of GR induced precession in Solar System chaos, in particular spin-orbit chaos, Mars' chaotic obliquity to be precise \citep[][]{tw93}. Indeed, in an ironic twist to Borges' observation, it is a case of an author being his/her own precursor, with \citep[][]{tw93} having noted how switching GR off in their simulations of Mars' obliquity in a fully evolving Solar System wiped off any sign of chaotic large amplitude excursions in Mars' spin axis. The authors take this as evidence of resonance detuning in view of the significant  contribution of GR to dominant secular orbital modes. They close their observation on a playful note: ``...Perhaps the geology of Mars will ultimately provide another test of the validity of general relativity..." [\cite[][]{tw93}], which we can liberally extend to our process asserting that the dearth of planets around binaries may well stand as yet another test of general relativity should more be called for in the age GW detections! Irony aside, the predominant trail on this end is largely structured around work on Mercury's predicament, and the salutary effect of GR precession in limiting the probability of its chaotic diffusion out of the Solar system by detuning secular resonances coupling the secular modes implicating Mercury and Jupiter primarily \citep[][]{laskar1994large,laskar2008chaotic,batygin2008dynamical,laskar2009existence,lithwick2011theory}. 
    
\end{itemize}
\section{By way of conclusion: A theme and its variations}\label{Section_Conclusion}

In this \textit{Letter}, motivated by the observed scarcity of transiting circumbinary  planets, we examined the effect of a secular resonance which obtains naturally in the course of the evolution of (near-)coplanar circumbinary systems. The resonance is encountered when the binary’s apsidal precession rate, set chiefly by general relativity, matches that of the planet which is in turn driven by the binary’s Newtonian perturbations. Capturing the system into this resonance transforms the otherwise reciprocal oscillatory  exchange of angular momentum between the binary and the planet into a unidirectional flow that steadily drains the planet’s angular momentum while exciting its eccentricity. We first located and characterized the resonance in phase space and outlined the dynamical outcomes expected upon capture. Next, we performed a suite of secular orbital simulations, with randomized initial conditions, to investigate how the secular resonance sculpts as it dictates the fate of a distribution of CBPs orbiting tidally shrinking binaries, and we learned the following:

\begin{itemize}
   \item[\textit{i)}] Systems which result in tight binaries (period $\leq7.45$ days, that of Kepler-47) via orbital decay, are more likely than not, deprived of a companion planet: the resonance-driven growth of the planet's eccentricity typically drives it into the throes of its host's driven instabilities, leading to ejection or engulfment by that host.

    \item[\textit{ii)}] Planetary survivors of the sweeping resonance mostly reside far from their host, and are therefore less likely to have their transits detected. Should eccentric survivors nevertheless be detected, they are expected to bear the signature of resonant capture into apse alignment with the binary.

    \item[\textit{iii)}] { The process appears robust to the modeling of the initial binary separation, with three out of four planets around tight binaries experiencing disruption, whether starting with Kepler-informed  or the more extreme log-uniform period distributions.}
   
\end{itemize}

Our results therefore suggest a significant role played by this secular resonance in shaping the architecture of circumbinary systems. In particular, we are proposing capture into a migrating apsidal GR-driven resonance as a natural and fairly robust explanation of the dearth of CBPs around tight binaries. { Moving forward, it would seem advisable to include GR-driven secular resonances in studies of the formation and dynamical evolution of planets around tightening binaries.}

An interesting byproduct of our experiments is the survival of a small family of planets on eccentric orbits around tight binaries. Taking the long-term stability of this relatively small population of planets for granted, one naturally wonders about the probability of them transiting the binary. Planetary eccentricity around single stars is known to enhance the transit probability by $(1-e_{\rm p}^2)^{-1}$ \citep[e.g.,][]{barnes2007effects}. But this effect has been obtained by marginalizing over all the possible apsidal orientations. CBPs that end up on stable eccentric orbits in our simulations are locked in the resonant apse-aligned configuration with the binary. To date, no study has thoroughly explored the effect of the eccentric orbits of both the planet and the binary when theoretically calculating transit probabilities. \citet[][]{martin2015circumbinary},  and later \citet[][]{martin2016circumbinary} performed numerical experiments that touched on these aspects --while mainly focused on the effect of mutual inclination-- but the effect of eccentricity remains partially quantified, and the existing results are inconclusive. Furthermore, while it can be reasonably expected, by analogy with single-star systems, that planetary eccentricity can increase the likelihood of transit, the binary's eccentricity and resonant capture into apse-alignment could partially work  against that enhancement. More specifically, while eccentricity brings the planet closer to the binary, apse-alignment forces this proximity to occur during the fastest passage of both the binary and the planet. Our results thus encourage future efforts in the direction of quantifying these effects.

While we have focused on the outcome of passage through resonance in the course of the binary's tidal orbital decay, we briefly touch on other evolutionary scenarios where the resonance is of potential significance. 

\begin{itemize}

\item \textit{Resonant transport with magnetic braking.} Rapidly rotating convective stars likely host strong magnetic dynamos. A magnetized wind co-rotates with the star out to the Alfv\'en radius \citep[][]{parker1958dynamics} and removes spin angular momentum \citep[][]{schatzman1962theory,mestel1987magnetic,reiners2012radius}. In tight binaries, stars are likely spin-orbit synchronized, and the wind-generated torque  depletes angular momentum from the orbital budget, rather than the spin, driving the binary to shrink \citep[e.g.,][]{ivanova2003magnetic,verbunt1981magnetic}. The resulting migration naturally drives encounters between CBPs and secular resonances tied to the binary, particularly the GR-driven apsidal resonance we study here. For example, using the braking prescription of \citet{verbunt1981magnetic}, we estimate that a solar-twin binary initialized at $P_{\rm AB;0}=2$~days has a characteristic orbital decay timescale of order $\sim3$~Gyr \citep[subject to the usual saturation scalings;][]{el2022magnetic}; ample time for adiabatic capture and excitation of  CBPs orbiting between ${\sim}10$ and ${\sim}80$ days. The further tightening of binaries under magnetic braking allows the resonance to operate disruptively on closer-in planets, potentially eroding some of the green survivors in Figure~\ref{Fig_final_eccentricities} that would not encounter the sweeping resonance during binary tidal decay.

Although our simulations focused on detached systems undergoing tidal evolution, semi-detached/contact binaries constitute 35\% of the Kepler eclipsing binaries catalogue, 80\% of which  have $P_{\rm AB}\leq1$~day. One can then envisage a two-stage evolution whereby such binaries start by decaying tidally from wider, eccentric progenitors until circularization on few-day periods, then continue their in-spiral via magnetic braking to produce tighter, sub-day periods. In this scenario, CBPs which were and remained captured in the resonance, acquiring sizable eccentricity by the end of the tidal decay phase (yellow family in Figure~\ref{Fig_final_eccentricities}), can experience further eccentricity growth toward near radial, hence destabilized, orbits during the braking phase. Magnetic braking would thus enhance the efficacy of the resonance as a CBP destruction channel, complementing the role played by tides.

\item \textit{Resonance encounter under planet migration.} There is of course the dual of the scenario considered here, namely that of resonance encounter in the course of a migrating planet around an effectively stationary binary (reasonable in view of the vast difference between the tidal evolution timescale and that of planetary migration). It is widely argued that most of the observed transiting planets did not form in situ \citep[e.g.,][]{paardekooper2012not,meschiari2014circumbinary,pierens2021vertical}, but rather further out in the proto-planetary disc and migrated inwards, eventually forced to `park' at their present locations by the density discontinuity at the inner edge of the disc \citep[e.g.,][]{pierens2008formation,thun2018migration,penzlin2021parking}.

Migration is rather treacherous, with the planets having to survive encounter with binary mean motion resonances mainly \citep[e.g.,][]{dvorak1984numerical,mardling2013new,sutherland2019instabilities,martin2022running,gianuzzi2023circumbinary}. Much has been explored on this front, but the effect of the secular precession resonance understudy is yet to be incorporated.

Evidently, the likelihood that a given planet encounters the GR induced resonance in the course of inward migration depends on its initial location in the disc. For the detected sample of transiting planets, we estimate the resonance location to range between ${\sim 1.7}~{\rm AU}$ for the compact Kepler-47 system, and ${\sim 7.3}~{\rm AU}$ in Kepler-34. As such, resonance encounter hinges on the planet's formation site before embarking on its inward migration journey. However, in contrast to the binary shrinkage scenario we examined in this work, a planet migrating inwards encounters the resonance  while crossing from $\dot{\varpi}_{\rm AB}>\dot{\varpi}_{\rm p}$ to $\dot{\varpi}_{\rm AB}<\dot{\varpi}_{\rm p}$. This crossing direction precludes resonance capture, but can certainly induce an eccentricity kick to the planet. Our preliminary explorations suggest that the amplitude of the kick is, to first order, and as should be expected, strongly dependent on the migration timescale. Fast Type-II migration (${\sim}10^3{-}10^4~$yr) allows the planet to traverse the resonant island in phase space without a noticeable change in angular momentum. Slower migration however (${\sim}10^5{-}10^6~$yr) can leave an initially circular planet with $e_{\rm p}{\sim}0.4{-}0.5.$ As such, in the latter regime, the survivability of the migrating planet hinges on  whether disc-driven eccentricity damping can counteract the resonant excitation before the orbit encroaches on the circumbinary instability zone. More interestingly, such an eccentricity excitation can, depending on the planetary mass and the disc's properties, and by virtue of torque reversal, halt the inward migration and even reverse its direction \citep[e.g.,][]{pierens2008evolution,bitsch2010orbital,pierens2013migration}. It is also noteworthy that, although inward migration is often presumed of the observed transiting CBPs, outward migration is also possible for massive planets ($m_{\rm p}\geq2~{M_{\rm J}})$ embedded in eccentric protoplanetary discs \citep[e.g.,][]{dempsey2021outward}. With disc asymmetries being naturally expected around binaries due to the binary' torques \citep[e.g.,][]{kley2008simulations,hirsh2020cavity}, outward migration offers another setting for capture into the secular resonance.

\item \textit{Resonance encounter under disk dispersal.} There is yet another, perhaps more promising, scenario for resonance migration, one which is shrouded in the early stage of planet formation: here neither planet nor binary orbits are evolving, rather the precession of both (the planet's primarily) is weakened in the course of proto-planetary disc dispersal. Indeed, the presence of a massive disc induces an additional precessional effect on the planet. As the gaseous component dissipates, the planet's precession slows down, allowing the system to transition from $\dot{\varpi}_{\rm AB}<\dot{\varpi}_{\rm p}$ to $\dot{\varpi}_{\rm AB}>\dot{\varpi}_{\rm p}$, i.e. in the `correct' direction of resonance capture. The understanding of disk dispersal around single, let alone binary stars, is work progress, but it is safe, between observation and theory, to expect discs to disperse over $\sim10^{5}-10^6,\mathrm{yr}$, far shorter than the binary’s tidal evolution, yet still slow enough for resonance encounter to be adiabatic, fulfilling a key condition for capture. This prospect leaves us in anticipation of a full fledged exploration of the early stages of coupled binary-planet-disc system. 
\end{itemize}

Needless to say, variations on coplanar scenarios with a single CBP can be further complexified, by going from coplanar to spatial on one hand, then going from single to multiplanets on the other, then revisiting the whole in S-type as opposed to the P-type systems we considered here. Variation in architecture and processes are currently under study by our team, with more to follow in the infamous class of future works (Al Kurdy, Farhat, Touma, In Preparation). 

\section*{Acknowledgments}
The authors thank the referee for a thorough review which improved the paper considerably. MF is supported by the Miller Institute for Basic Research in Science at UC Berkeley through a Miller Research Fellowship. MF expresses his gratitude to Eugene Chiang and Amaury Triaud for helpful discussions. MF and JT thank Aya Al Kurdy whose careful probing spared us a glitch in test particle resonance tracks. JT would like to recognize early insights on the process from Petra Awad's undergraduate research project (2017-2019) on planet formation around pulsar binaries. 
\newpage
\appendix
\section*{secular Equations governing the dynamical evolution of a circumbinary planet}
\label{Appendix_EQM}
\textbf{---The Secular Hamiltonian:} We consider a stellar binary with masses $m_{\rm A}$ and $m_{\rm B}$, on an orbit defined by the semi-major axis $a_{\rm AB}$, eccentricity $e_{\rm AB}$, angular momentum $L_{\rm AB}=(m_{\rm A}m_{\rm B}/m_{\rm AB})\sqrt{Gm_{\rm AB}a_{\rm AB}(1-e_{\rm AB}^2)}$, and the vectors $\vec{j}_{\rm AB}=\sqrt{1-e_{\rm AB}^2}\,\hat{n}_{\rm AB}$ and $\vec{e}_{\rm AB}$ such that $\vec{j}_{\rm AB}^{\,2}+ \vec{e}_{\rm AB}^{\,2}=1$, and $\hat{n}_{\rm AB}$ defines the direction of the binary's orbital angular momentum vector. A circumbinary planet of mass $m_{\rm p}$ orbits this inner binary in a hierarchical configuration with the analogous elements $a_{\rm p},e_{\rm p},\vec{j}_{\rm p},\vec{e}_{\rm p},$ and $L_{\rm p}=(m_{\rm AB}m_{\rm p}/m_{\rm ABp})\sqrt{Gm_{\rm ABp}a_{\rm p}(1-e_{\rm p}^2)}$, with $m_{\rm ABp}=m_{\rm AB}+m_{\rm p}$. 

The secular Hamiltonian describing the dynamical evolution of the triple is given by
\begin{align}\nonumber
      H_{\rm S}&= H_{\rm K;AB} +H_{\rm K;p} + H_{\rm N} + H_{\rm GR; AB} \\
      &=-\frac{Gm_{\rm A}m_{\rm B}}{2a_{\rm AB}}-\frac{Gm_{\rm p}m_{\rm AB}}{2a_{\rm p}}+\frac{Gm_{\rm p}m_{\rm AB}}{a_{\rm p}}\left(\Psi_{\rm N,quad}+\Psi_{\rm N,oct}\right)+H_{\rm GR; AB}.\label{Ham_1}
\end{align}
Here, the first two components correspond to the constant Keplerian parts of the total Hamiltonian; $H_{\rm N}$ describes the Newtonian gravitational interaction between the binary and the planet, obtained by computing the secularly averaged disturbing function $\Psi_{\rm N}$ up to the octupolar order; and  $H_{\rm GR; AB}$ delivers the dominant secular contribution of the general relativistic effect in the binary. Specifically:
\begin{equation}
    \Psi_{\rm N,quad}=\frac{3}{8}\frac{\varepsilon_{\rm q}}{(1-e_{\rm p}^2)^{5/2}}\left[ 5(\vec j_{\rm p}\cdot \vec e_{\rm AB})^2-(1-e_{\rm AB}^2)(\vec j_{\rm p}\cdot \hat{n}_{\rm AB})^2 +(1-e_{\rm p}^2)(1/3 -2e_{\rm AB}^2)\right],
\label{Psi_N_quad}
\end{equation}
\begin{align}\nonumber
     \Psi_{\rm N,oct}=&\frac{15}{64}\frac{\varepsilon_{\rm q}\varepsilon_{\rm o}}{(1-e_{\rm p}^2)^{7/2}} \Big\{ (\vec e_{\rm p}\cdot \vec e_{\rm AB})\left[(8e_{\rm AB}^2-1)(1-e_{\rm p}^2)-35(\vec j_{\rm p}\cdot \vec e_{\rm AB})^2+5(1-e_{\rm AB}^2)(\vec j_{\rm p}\cdot \hat{n}_{\rm AB})^2\right]
     \\&+10(1-e_{\rm AB}^2)(\vec j_{\rm p}\cdot \vec e_{\rm AB})(\vec e_{\rm p}\cdot \hat{n}_{\rm AB})(\vec j_{\rm p}\cdot \hat{n}_{\rm AB}) \Big\},
\label{Psi_N_oct}
\end{align}
and
\begin{equation}
    H_{\rm GR; AB} = -\frac{3G^2m_{\rm A}m_{\rm B}m_{\rm AB}}{c^2 a_{\rm AB}^2\sqrt{1-e_{\rm AB}^2}}\,,
\end{equation}
where we have defined the dimensionless quadrupolar coefficient $\varepsilon_{\rm q}=\alpha^2 {m_{\rm A}m_{\rm B}}/{m_{\rm AB}^2},$ and the octupolar coefficient $\varepsilon_{\rm o}=\alpha({m_{\rm A}-m_{\rm B}})/{m_{\rm AB}},$ where $\alpha=a_{\rm AB}/a_{\rm p}.$ If one restricts the system to the coplanar setting, the Newtonian interaction between the planet and the binary drives apsidal precession in the stellar binary of the form:
\begin{equation}
    \dot\omega_{\rm AB;N} =\frac{\varepsilon_{\rm q}}{8L_{\rm AB}}\frac{Gm_{\rm AB}m_{\rm p}}{a_{\rm p}}\frac{\sqrt{1-e_{\rm AB}^2}}{(1-e_{\rm p}^2)^{3/2}} \left[6 - \frac{15\varepsilon_{\rm o}}{8}\frac{e_{\rm p}}{1-e_{\rm p}^2}\left(9e_{\rm AB} + \frac{4}{e_{\rm AB}}\right) \cos\Delta\varpi \right],
\end{equation}
and in the planet's orbit in the form:
\begin{equation}
     \dot\omega_{\rm p;N} = \frac{3}{4L_{\rm p}}\frac{Gm_{\rm AB}m_{\rm p}}{a_{\rm p}}\frac{\varepsilon_{\rm q}}{(1-e_{\rm p}^2)^2} \left[\left(1+\frac{3}{2}e_{\rm AB}^2\right) - \frac{5\varepsilon_{\rm o}}{8} e_{\rm AB}\left(2+3/2 e_{\rm AB}^2\right) \frac{1+4e_{\rm p}^2}{e_{\rm p}(1-e_{\rm p}^2)} \cos\Delta\varpi\right].
\end{equation}
The general relativistic contribution delivered by $H_{\rm GR; AB}$ forces the binary to precess following:
\begin{equation} 
\dot{\omega}_{\rm GR}=\frac{3Gm_{\rm AB}}{c^2 a_{\rm AB}} \frac{n_{\rm AB}}{1-e_{\rm AB}^2}.
\end{equation}
The Hamiltonian of \eq{Hamiltonian_1} governing the dynamics of a circumbinary test particle, and generating the phase space structure of Figure \ref{Figure_CBP_phase_space}, is obtained by restricting the Hamiltonian of \eq{Ham_1} to the planar configuration, assuming a test particle approximation ($m_{\rm p}\rightarrow0)$, and transforming to the frame co-rotating with the precessing binary under GR. The transformation involves adding a rotational contribution of the form:
\begin{equation}
    H_{\rm rot} = -\dot{\omega}_{\rm GR} \sqrt{G m_{\rm AB}a_{\rm p} }\,(\vec j_{\rm p}\cdot\hat n_{\rm AB}). 
\end{equation}
\textbf{---The Governing Equations of Motion:} With $H_{\rm S}$ fully defined in the general form of \eq{Ham_1}, we derive the equations of motion governing the dynamical evolution of the system in the secular, orbit-averaged regime, whereby the orientation and shape of the orbits are described by the vectors $\vec{j}_{\rm p},\vec{e}_{\rm p},\vec{j}_{\rm AB},$ and $\vec{e}_{\rm AB}$, preserving the semi-major axes $a_{\rm p}$ and $a_{\rm AB}$, unless the latter are forced to evolve by additional mechanisms. In this vectorial description, the equations of motion take the form \citep[e.g.,][]{milankovitch1939verwendung,allan1964long,tremaine2009satellite}:
\begin{align}
    \frac{d\vec{j}_{\rm AB}}{dt} &= -\frac{1}{L_{\rm AB}}\left(\vec{j}_{\rm AB}\times\nabla_{\vec{j}_{\rm AB}}H_{\rm S} + \vec{e}_{\rm AB}\times \nabla_{\vec{e}_{\rm AB}}H_{\rm S}\right),\\
     \frac{d\vec{e}_{\rm AB}}{dt} &= -\frac{1}{L_{\rm AB}}\left(\vec{j}_{\rm AB}\times\nabla_{\vec{e}_{\rm AB}}H_{\rm S} + \vec{e}_{\rm AB}\times \nabla_{\vec{j}_{\rm AB}}H_{\rm S}\right),\\
      \frac{d\vec{j}_{\rm p}}{dt} &= -\frac{1}{L_{\rm p}}\left(\vec{j}_{\rm p}\times\nabla_{\vec{j}_{\rm p}}H_{\rm S} + \vec{e}_{\rm p}\times \nabla_{\vec{e}_{\rm p}}H_{\rm S}\right),\\
       \frac{d\vec{e}_{\rm p}}{dt} &= -\frac{1}{L}_{\rm p}\left(\vec{j}_{\rm p}\times\nabla_{\vec{e}_{\rm p}}H_{\rm S} + \vec{e}_{\rm p}\times \nabla_{\vec{j}_{\rm p}}H_{\rm S}\right).
\end{align}
For the Newtonian contribution we obtain:
\begin{align}\nonumber
    \frac{d\vec j_{\rm p}}{dt}^{\rm (N)}&= \frac{3}{4}\varepsilon_{\rm q} {n}_{\rm p}(1-e_{\rm p}^2)^{-5/2}\left[(\vec j_{\rm p} \cdot \vec{j}_{\rm AB})(\vec j_{\rm p} \times \vec{j}_{\rm AB}) -5(\vec j_{\rm p} \cdot\vec e_{\rm AB})(\vec j_{\rm p} \times\vec e_{\rm AB})\right] \\
    &-\frac{15}{64}\varepsilon_{\rm q}\varepsilon_{\rm o}{n}_{\rm p}{(1-e_{\rm p}^2)^{-7/2}}\left[
    (\Psi_1 \vec e_{\rm p} +\Psi_2 \vec j_{\rm p}) \times \vec e_{\rm AB} + (\Psi_3 \vec e_{\rm p} + \Psi_4 \vec j_{\rm p}) \times  \vec{j}_{\rm AB} \right],\label{djdt_N}
\end{align}
\begin{align}\nonumber
    \frac{d\vec e_{\rm p}^{\,\rm (N)}}{dt}=& \frac{3}{4}\varepsilon_{\rm q} n_{\rm p}\Bigg\{{(1-e_{\rm p}^2)^{-5/2}}\Big((\vec j_{\rm p} \cdot \vec{j}_{\rm AB})(\vec e_{\rm p} \times \vec{j}_{\rm AB}) -5(\vec j_{\rm p} \cdot\vec e_{\rm AB})(\vec e_{\rm p} \times\vec e_{\rm AB})\Big)\\\nonumber
    &-\frac{1}{2}(1-e_{\rm p}^2)^{-7/2}\Big((1-6e_{\rm AB}^2)(1-e_{\rm p}^2)-5(\vec j_{\rm p} \cdot \vec{j}_{\rm AB})^2 +25(\vec j_{\rm p} \cdot\vec e_{\rm AB})^2\Big)\vec j_{\rm p}\times \vec e_{\rm p}\Bigg\}\\\nonumber
 &-\frac{15}{64}\varepsilon_{\rm q}\varepsilon_{\rm o} n_{\rm p}\Bigg\{
    (1-e_{\rm p}^2)^{-7/2}\Big((\Psi_1 \vec j_{\rm p} +\Psi_2 \vec e_{\rm p}) \times \vec e_{\rm AB} + (\Psi_3 \vec j_{\rm p} + \Psi_4 \vec e_{\rm p}) \times  \vec{j}_{\rm AB})\Big)\\
    &+(1-e_{\rm p}^2)^{-9/2}\Big(\Psi_5 +\Psi_2(\vec j_{\rm p} \cdot \vec e_{\rm AB}) +\Psi_4(\vec j_{\rm p} \cdot \vec{j}_{\rm AB})\Big)\vec j_{\rm p} \times \vec e_{\rm p}\Bigg\},\label{dedt_N}
\end{align}
\begin{align}
     \frac{d\vec j_{\rm AB}}{dt}^{\rm (N)}&= -\frac{L_{\rm p}}{L_{\rm AB}} \frac{d\vec j_{\rm p}}{dt}^{\rm (N)},
\end{align}
\begin{align}\nonumber
    \frac{d\vec e^{\,\rm (N)}_{\rm AB}}{dt}=&\frac{3}{4}\frac{L_{\rm p}}{L_{\rm AB}}\varepsilon_{\rm q} n_{\rm p}(1-e_{\rm p}^2)^{-5/2} \left[5(\vec j_{\rm p}\cdot \vec e_{\rm AB})(\vec j_{\rm p}\times \vec j_{\rm AB}) + 2(1-e_{\rm p}^2)(\vec j_{\rm AB}\times \vec e_{\rm AB})  - (\vec j_{\rm p}\cdot \vec j_{\rm AB})(\vec j_{\rm p}\times \vec e_{\rm AB})\right]\\
    & + \frac{15}{64}\frac{L_{\rm p}}{L_{\rm AB}}\varepsilon_{\rm q}\varepsilon_{\rm o} n_{\rm p}(1-e_{\rm p}^2)^{-7/2}\left[\left(\Psi_1\vec e_{\rm p} + \Psi_2 \vec j_{\rm p} + \Psi_6\vec e_{\rm AB}\right)\times \vec j_{\rm AB} + \left(\Psi_3 \vec e_{\rm p} + \Psi_4\vec j_{\rm p}\right)\times \vec e_{\rm AB}\right].
\end{align}
Here we have defined the following functions:
\begin{align}
    \nonumber
   & \Psi_1=  8e_{\rm AB}^2(1-e_{\rm p}^2) -1+e_{\rm p}^2 +5(\vec j_{\rm p} \cdot \vec{j}_{\rm AB})^2 -35(\vec j_{\rm p} \cdot \vec e_{\rm AB})^2\\    \nonumber
   & \Psi_2=10(\vec e_{\rm p}\cdot \vec{j}_{\rm AB})(\vec j_{\rm p} \cdot \vec{j}_{\rm AB}) -70(\vec j_{\rm p} \cdot \vec e_{\rm AB})(\vec e_{\rm p} \cdot \vec e_{\rm AB})\\    \nonumber
   & \Psi_3=10(\vec j_{\rm p} \cdot \vec e_{\rm AB})(\vec j_{\rm p} \cdot \vec{j}_{\rm AB})\\    \nonumber
   &\Psi_4=10\Big[(\vec e_{\rm p} \cdot \vec e_{\rm AB})(\vec j_{\rm p} \cdot \vec{j}_{\rm AB}) +(\vec j_{\rm p} \cdot \vec e_{\rm AB})(\vec e_{\rm p} \cdot \vec{j}_{\rm AB})\Big]\\\nonumber
   &\Psi_5=5\Big[\Psi_1(\vec e_{\rm p} \cdot \vec e_{\rm AB}) +\Psi_3 (\vec e_{\rm p}\cdot \vec{j}_{\rm AB})\Big]\\
   &\Psi_6 =16(\vec e_{\rm p}\cdot\vec e_{\rm AB})(1-e_{\rm p}^2).
\end{align}
As for the GR contribution, we obtain
\begin{equation}
 \frac{d\vec j_{\rm AB}}{dt}^{\rm (GR)}=0,
\end{equation}
\begin{equation}
 \frac{d\vec e_{\rm AB}}{dt}^{\rm (GR)}=\frac{3G^{3/2}m_{\rm AB}^{3/2}}{c^2 a_{\rm AB}^{5/2}(1-e^2_{\rm AB})^{3/2}}\,\vec j_{\rm AB}\times \vec e_{\rm AB},
\end{equation}
\begin{equation}
 \frac{d\vec j_{\rm p}}{dt}^{\rm (GR)}=0,
\end{equation}
\begin{equation}
 \frac{d\vec e_{\rm p}}{dt}^{\rm (GR)}=\frac{3G^{3/2}m_{\rm ABp}^{3/2}}{c^2 a_{\rm p}^{5/2}(1-e_{\rm p}^2)^{3/2}}\,\vec j_{\rm p}\times \vec e_{\rm p}.
\end{equation}
\textbf{---Equilibrium Conditions:} To obtain the equilibrium conditions of these dynamical equations, we set $d\bold{j}_{\rm p}/dt = d\bold{e}_{\rm p}/dt =0$, and we restrict the three-dimensional equations to a coplanar-coplanar configuration, in which the vectors $\bold{j}_{\rm p}, {e}_{\rm p}, \bold{j}_{\rm AB}$, and $\bold{e}_{\rm AB}$ live in the same plane, i.e. aligning the lines of nodes of the CBP with that of the binary. We therefore recover two scalar equations with two unknowns: the CBP's eccentricity, $e_{\rm p}$, and its inclination $i_{\rm p}$ with respect to the binary. These conditions read as:
\begin{align}
\nonumber
    0&= \frac{3\varepsilon_{\rm q}}{8}\frac{4e_{\rm AB}^2+1}{(1-e_{\rm p}^2)^\frac{3}{2}} \sin{2i_{\rm p}}-\frac{15\varepsilon_{\rm q} \varepsilon_{\rm o}}{64}\frac{e_{\rm p}e_{\rm AB}}{(1-e_{\rm p}^2)^\frac{5}{2}}\times\Big(\cos^2i_{\rm p}\left(60e_{\rm AB}^2+45\right)-17e_{\rm AB}^2-11 \Big) \sin i_{\rm p}\cos\Delta\varpi \\
    & +\dot\varpi_{\rm GR }\sqrt{1-e_{\rm p}^2}\sin i_{\rm p}, \label{equilibria_dj}
\end{align}
\begin{align}
\nonumber
    0&= \frac{9\varepsilon_{\rm q}}{8}\frac{ e_{\rm p}}{(1-e_{\rm p}^2)^2}\left[-\sin^2{i_{\rm p}}(4e_{\rm AB}^2+1)+e_{\rm AB}^2+2/3\right]-\frac{15\varepsilon_{\rm q} \varepsilon_{\rm o}}{64}\frac{e_{\rm AB} (4e_{\rm p}^2+1)}{(1-e_{\rm p}^2)^3}\cos i_{\rm p}\cos\Delta\varpi\times \\
    &\Big(\cos^2 i_{\rm p}(20e_{\rm AB}^2+15) -17e_{\rm AB}^2-11\Big) -\dot\varpi_{\rm GR }e_{\rm p}\cos i_{\rm p} \cos \Delta\varpi.\label{equilibria_de}
\end{align}
\textbf{---The Effects of Gravitational Waves Emission:} Dissipation in the stellar binary due to gravitational waves emission draws energy and angular momentum from the orbit following:
\begin{equation}
     \frac{d\vec j_{\rm AB}^{\rm (GW)}}{dt} = -\frac{32}{5}\frac{G^3m_{\rm AB}m_{\rm A}m_{\rm B}}{c^5 a_{\rm AB}^4}\frac{1+7e_{\rm AB}^2/8}{(1-e_{\rm AB}^2)^{5/2}}\,\vec j_{\rm AB},
\end{equation}
\begin{equation}
    \frac{d\vec e_{\rm AB}^{\rm\, (GW)}}{dt} = -\frac{304}{15}\frac{G^3m_{\rm AB}m_{\rm A}m_{\rm B}}{c^5 a_{\rm AB}^4}\frac{1+121e_{\rm AB}^2/304}{(1-e_{\rm AB}^2)^{5/2}}\,\vec e_{\rm AB},
\end{equation}
\begin{equation}
    \frac{da_{\rm AB}^{\rm\, (GW)}}{dt}= -\frac{64}{5}\frac{G^3m_{\rm AB}m_{\rm A}m_{\rm B}}{c^5 a_{\rm AB}^3}\frac{1+73e_{\rm AB}^2/24 + 37e_{\rm AB}^4/96}{(1-e_{\rm AB}^2)^{7/2}}\,.
\end{equation}
\textbf{---The Effects of Tides:} What is left to complete the system studied in Section \ref{Section_tidal_evolution} is to add the effect of tidal interactions within the binary and between the binary and the planet. We therefore ignore tides raised on the stars by the planet. We adopt a weak friction tidal model with a constant time lag $\Delta t$ to describe the tidal responses of each of the three bodies \citep[e.g.,][]{mignard1979evolution,hut1980stability}. Similar to the gravitational potential above, we average the tidal potential over the mean anomalies of the binary and the stars to obtain the secular equations of motion. We further assume spin-orbit alignment for the stellar binary, and a pseudo-synchronized equilibrium rotation (i.e., $\Omega_{\rm A} = \Omega_{\rm B} = n_{\rm AB}f_2(e_{\rm AB)}/f_1 (e_{\rm AB})$). Under these assumptions, the evolution of the binary's Lenz vector is described as:  
\begin{align}\nonumber
  \frac{d\vec e_{\rm AB}}{dt}^{\rm (T)}&= \frac{15}{2}n_{\rm AB} \frac{f_4(e_{\rm AB)}}{(1-e_{\rm AB}^2)^{1/2}}\left[ \frac{m_{\rm B}}{m_{\rm A}}\left( \frac{R_{\rm A}}{a_{\rm AB}}\right)^5 k_{2,\rm A} +\frac{m_{\rm A}}{m_{\rm B}}\left( \frac{R_{\rm B}}{a_{\rm AB}}\right)^5 k_{2,\rm B} \right]\vec{j}_{\rm AB}\times \vec{e}_{\rm AB} \\
 & -3G \frac{m_{\rm AB}}{m_{\rm A}m_{\rm B}}\frac{\Delta t_{\rm AB}}{a_{\rm AB}^8}\left[ 9f_5(e_{\rm AB})-\frac{11}{2}\frac{f_4(e_{\rm AB}) f_2(e_{\rm AB})}{f_1(e_{\rm AB})} \right]\left(m_{\rm B}^2 R_{\rm A}^5 k_{2,\rm A} + m_{\rm A}^2 R_{\rm B}^5 k_{2,\rm B} \right) \vec{e}_{\rm AB},\label{Eq_tidal_eAB}
\end{align}
where $k_{2,i}$ is the second harmonic gravitational Love number for each star, and $\Delta t_{\rm AB}$ is the time deformation time lag of each star. In our simulations we adopt $k_{2,A}=k_{2,\rm B}=0.028$ \citep[][]{eggleton2001orbital}, and we sample $\Delta t_{\rm AB}$ in the population synthesis from a uniform distribution between $0.1$~s and 10~s, implying a range for the stellar quality factor $Q{\sim}10^6{-}10^8$, which is consistent with analyses of observed stellar spins, orbital circularization timescales of binaries, and orbital configurations of close-in exoplanets \citep[e.g.,][]{meibom2006observational,penev2012constraining,patel2022constraining}. The eccentricity functions used in \eq{Eq_tidal_eAB} read as:
\begin{align}
f_{1}(e) &= \frac{1 + 3e^2 + \frac{3}{8}e^4}{\bigl(1 - e^2\bigr)^{9/2}}, \\
f_{2}(e) &= \frac{1 + \tfrac{15}{2}e^2 + \tfrac{45}{8}e^4 + \tfrac{5}{16}e^6}{\bigl(1 - e^2\bigr)^{6}}, \\
f_{3}(e) &= \frac{1 + \tfrac{31}{2}e^2 + \tfrac{255}{8}e^4 + \tfrac{185}{16}e^6 + \tfrac{25}{64}e^8}{\bigl(1 - e^2\bigr)^{15/2}},\\
f_{4}(e) &= \frac{1 + \tfrac{3}{2}e^2 + \tfrac{1}{8}e^4}{\bigl(1 - e^2\bigr)^{5}},\\
f_{5}(e) &= \frac{1 + \tfrac{15}{4}e^2 + \tfrac{15}{8}e^4 + \tfrac{5}{64}e^6}{\bigl(1 - e^2\bigr)^{13/2}}.
\end{align}
{The tidal torque does not force $\vec j_{\rm AB}$ to precess, but it does change its magnitude in lockstep with $\vec e_{\rm AB}$ to maintain $|\vec j_{\rm AB}|^2 +|\vec e_{\rm AB}|^2=1;$ thus
\begin{equation}
   \frac{d\vec j_{\rm AB}^{(\rm T)}}{dt}  = -\frac{e_{\rm AB}\dot{e}_{\rm AB}}{|\vec j_{\rm AB}|^2}\vec j_{\rm AB},
\end{equation}
where $\dot{e}_{\rm AB} = (\vec{e}_{\rm AB}\cdot \dot{\vec{e}}_{\rm AB})/{e}_{\rm AB} = D e_{\rm AB},$ $D$ being the damping component (the negative, second term) of \eq{Eq_tidal_eAB}.
Tidal dissipation in the binary induces orbital decay according to:
\begin{equation}
   \frac{\dot{a}_{\rm AB}^{\rm T}}{a_{\rm AB}} =\frac{2e_{\rm AB}\dot{e}_{\rm AB}}{1-e_{\rm AB}^2}.
\end{equation}}



\end{document}